\documentclass[twocolumn,tighten,times]{aastex631}

\usepackage{ragged2e}

\newcommand{\um}{$\mu$m}
\def\m51{{M\nolinebreak[4]\hspace{0.08em}51}}

\shorttitle{B-field structure in spiral galaxies}
\shortauthors{Surgent, Lopez-Rodriguez, \& Clark}
\begin{document}

\title{The structure of magnetic fields in spiral galaxies: a radio and far-infrared polarimetric analysis}

%\correspondingauthor{William Jeffrey Surgent \& Enrique Lopez-Rodriguez}
%\email{wsurgent@stanford.edu} \email{elopezrodriguez@stanford.edu}

\author[0000-0002-8108-6904]{William Jeffrey Surgent}
\affiliation{Department of Physics, Stanford University, Stanford, California 94305, USA}
\affiliation{Kavli Institute for Particle Astrophysics \& Cosmology (KIPAC), Stanford University, Stanford, CA 94305, USA}

\author[0000-0001-5357-6538]{Enrique Lopez-Rodriguez}
\affiliation{Kavli Institute for Particle Astrophysics \& Cosmology (KIPAC), Stanford University, Stanford, CA 94305, USA}

\author[0000-0002-7633-3376]{Susan E. Clark}
\affiliation{Department of Physics, Stanford University, Stanford, California 94305, USA}
\affiliation{Kavli Institute for Particle Astrophysics \& Cosmology (KIPAC), Stanford University, Stanford, CA 94305, USA}

\begin{abstract} 
We propose and apply a method to quantify the morphology of the large-scale ordered magnetic fields (B-fields) in galaxies. This method is adapted from the analysis of Event Horizon Telescope polarization data. We compute a linear decomposition of the azimuthal modes of the polarization field in radial galactocentric bins. We apply this approach to five low-inclination spiral galaxies with both far-infrared (FIR: $154$ \um) dust polarimetric observations taken from the Survey of ExtragALactic magnetiSm with SOFIA (SALSA) and radio ($6$ cm) synchrotron polarization observations. We find that the main contribution to the B-field structure of these spiral galaxies comes from the $m=2$ and $m=0$ modes at FIR wavelengths and the $m=2$ mode at radio wavelengths. The $m=2$ mode has a spiral structure and is directly related to the magnetic pitch angle, while $m=0$ has a constant B-field orientation. The FIR data tend to have a higher relative contribution from other modes than the radio data. The extreme case is NGC~6946: all modes contribute similarly in the FIR, while $m=2$ still dominates in the radio. The average magnetic pitch angle in the FIR data is smaller and has greater angular dispersion than in the radio, indicating that the B-fields in the disk midplane traced by FIR dust polarization are more tightly wound and more chaotic than the B-field structure in the radio, which probes a larger volume. We argue that our approach is more flexible and model-independent than standard techniques, while still producing consistent results where directly comparable.
\end{abstract}

%% Keywords should appear after the \end{abstract} command. 
%% The AAS Journals now uses Unified Astronomy Thesaurus concepts:
%% https://astrothesaurus.org
%% You will be asked to selected these concepts during the submission process
%% but this old "keyword" functionality is maintained in case authors want
%% to include these concepts in their preprints.

%% From the front matter, we move on to the body of the paper.
%% Sections are demarcated by \section and \subsection, respectively.
%% Observe the use of the LaTeX \label
%% command after the \subsection to give a symbolic KEY to the
%% subsection for cross-referencing in a \ref command.
%% You can use LaTeX's \ref and \label commands to keep track of
%% cross-references to sections, equations, tables, and figures.
%% That way, if you change the order of any elements, LaTeX will
%% automatically renumber them.
%%
%% We recommend that authors also use the natbib \citep
%% and \citet commands to identify citations.  The citations are
%% tied to the reference list via symbolic KEYs. The KEY corresponds
%% to the KEY in the \bibitem in the reference list below. 

%%%%%%%%%%%%%%%%%%%%%%%%%%%%%%%%%%%%%%%%%%%%%%%%%%%%%%%%
\section{Introduction} \label{sec:INT}
%%%%%%%%%%%%%%%%%%%%%%%%%%%%%%%%%%%%%%%%%%%%%%%%%%%%%%%%

Large-scale spiral magnetic field (B-field) structures are frequently observed in spiral galaxies \citep[e.g.,][]{Beck2019,SALSAVI}. 
These B-fields are thought to be generated via a mean-field dynamo driven by differential rotation of the galactic disk and turbulent helical motions \citep{ss21}. The three-dimensional structure of galactic B-fields can be decomposed into radial ($B_{\rm r}$), azimuthal ($B_{\phi}$), and vertical ($B_{\rm z}$) components, where the coordinate system is typically defined relative to the core of the galaxy. The structure of the disk magnetic field (e.g., at the midplane $z = 0$) is often summarized using the pitch angle $\Psi_{\rm B}=\arctan(B_{\rm r}/B_{\rm \phi})$ \citep{Krasheninnikova1989}. In this formalism, a perfectly toroidal B-field has $\Psi_{\rm B} =0^{\circ}$, and a perfectly radial B-field has $\Psi_{\rm B} = 90^{\circ}$. 

The full three-dimensional structure of galactic B-fields is not directly measurable, but $\Psi_{\rm B}$ can be estimated from polarimetric measurements of a galactic disk. For any point in the galaxy, $\Psi_{\rm B}$ is the angle between the local magnetic field orientation and the tangent to a circle with origin at the galaxy's center that passes through that point. The latest compilation of $\Psi_{\rm B}$ from radio polarimetric observations of $19$ nearby galaxies shows that the $\Psi_{\rm B}$ is mostly constant within the central $5-10$ kpc, with values in the range of $20-35^{\circ}$ \citep{Beck2019}. The $\Psi_{\rm B}$ were found to be systematically offset by $5–10^{\circ}$ when compared with the molecular (CO) spiral arms \citep{VanEck2015,Frick2016}---i.e, the magnetic pitch angles are more open than the molecular gas arms.

Far-infrared (FIR) polarimetric observations have shown to reveal different components of the large-scale B-fields in the disk of galaxies \citep[e.g.,][]{SALSAI,SALSAII,SALSAVI,SALSAV}. The detailed study performed in the spiral galaxy M51 showed that the radio and FIR magnetic pitch angles are similar within the central $6$ kpc, but at larger radii the FIR $\Psi_{\rm B}$ wrapped tighter than at radio wavelengths \citep{SALSAI}. The reason for this difference may be caused by the interaction of M51 with M51b (NGC~5195) and/or the injection of kinetic energy driven by the strong star formation region in the outskirts of the spiral arms of M51. These results have been further confirmed using a sample of seven spiral galaxies with FIR and radio polarimetric observations \citep{SALSAV}. The difference in the observed B-field structure arises from the different nature of the interstellar medium (ISM) associated with the FIR and radio wavelengths. The FIR polarization arises from thermal emission of magnetically aligned dust grains tracing a density-weighted medium along the line-of-sight (LOS) and within the beam (i.e., full-width-at-half-maximum, FWHM) of the observations of a dense ($\log_{10}(N_{\rm{HI+H2}} [\rm{cm}^{-2}]) = [19.96, 22.91]$) and cold ($T_{\rm d} = [19, 48]$ K) component of the ISM \citep[SALSA~IV, ][]{SALSAIV}. The radio polarimetric observations arise from non-thermal synchrotron emission in the warm and diffuse ISM. The radio polarized intensity has been found to be higher in the interarm regions and halos of galaxies \citep[e.g.,][]{Beck2019,Krause2020}.

The pitch angles have been characterized assuming a priori functions meant to represent spiral arms---i.e., logarithmic spirals. Some models estimate pitch angles using a logarithmic spiral function fitted to the spiral arms of the gas tracers and/or a multi-mode logarithmic spiral B-field fitted to the magnetic arms \citep{Fletcher2011,VanEck2015}. Then, a mean pitch angle value of the entire galaxy is estimated and compared between tracers. This approach assumes that the B-field is well-described by a single spiral B-field. 

Other models used a wavelet-based approach in the spiral galaxies M51 \citep{Patrikeev2006}, M83 \citep{Frick2016}, and NGC~6946 \citep{Frick2000}. \citet{SALSAI} performed the same wavelet analysis to the morphological spiral arms and estimated the mean pitch angle per annulus as a function of galactocentric radius for the magnetic pitch angles. All these studies found an almost flat magnetic pitch angle as a function of the galactocentric radius, and systematic angular offsets between the magnetic pitch angle and the gas structures (i.e., CO, HI, and HII emission). These approaches use a kernel with a certain profile, typically some class of wavelet, to estimate the width and orientation of the spatial structures in polarized emission. The output of this method is highly dependent on the applied kernel.
%The output also depends on the width of the kernel, which is typically selected based on the FWHM of the observations and the width of the structure that is characterized.

Other works have analyzed the regular B-field structure of galaxies using linear models of the mean-field galactic dynamo \citep[e.g.,][]{KW1991,Berkhuijsen1997,Fletcher2004,Fletcher2011}. These models assume an expanded B-field pattern in a Fourier series in the azimuthal angle. Each mode is a logarithmic spiral with a constant magnetic pitch angle, with the sum of all modes describing a non-axisymmetric B-field. The radio B-field orientations, corrected for Faraday rotation, are fitted using a linear superposition of logarithmic spiral B-fields in three dimensions. The azimuthal wave number $m_{\rm{d}} = 0$ is an axisymmetric B-field with constant B-field direction, $m_{\rm{d}} = 1$ is a bisymmetric B-field with two opposite spiral B-field directions, and $m_{\rm{d}} = 2$ is a quadrisymmetric B-field with alternating B-field directions. Most of the studied galaxies in radio polarimetric observations are dominated by $m_{\rm{d}}=0$ \citep{Beck2019}, although higher modes are sometimes required \citep[e.g.,][]{Ehle1993,Rohde1999}. $m_{\rm{d}}>2$ cannot be studied with the spatial resolutions provided by current radio polarimetric observations because they are not sensitive to small spatial variations of the B-field direction. For some galaxies, any combination of modes provides a good fit for the B-field orientations \citep[][table 6]{Beck2019}. Although a decomposition of the B-field is made, only linear combination of logarithmic spirals are taken into account. Thus, a model-independent systematic study of the B-field geometry as a function of galactocentric radii and tracers is required.

Our goal is to characterize the B-field morphology of spiral galaxies. We make use of the linear polarimetric decomposition approach \citep{Palumbo2020} applied to analyze the B-field structure around the supermassive black hole of M87 from the Event Horizon Telescope \citep[EHT;][]{EHT2021_VII,EHT2021_VIII}. \citet[][SALSA~II]{SALSAII} applied this method to the B-fields in the starburst ring of NGC~1097, finding that the radio B-field is dominated by a spiral B-field (with an azimuthal mode $m=2$), while a constant ($m=0$) B-field dominates at FIR wavelengths. The $m=2$ B-field was attributed to a magnetohydrodynamic (MHD) dynamo, and the $m=0$ B-field was associated with galactic shocks between the bar and the starburst ring. These results showed the potential of this method to analyze the B-fields in the multiphase of galaxies. Here, we apply the linear polarimetric decomposition to analyze the B-field orientation in a sample of five spiral galaxies with resolved radio and FIR polarimetric observations. We describe in Section \ref{sec:MET} the methodology of the linear polarimetric decomposition. The results of the decomposition of the B-field morphology using FIR and radio observations are shown in Section \ref{sec:APPLICATION}. Our discussions are described in Section \ref{sec:DIS}, and our main conclusions are summarized in Section \ref{sec:CON}.

%%%%%%%%%%%%%%%%%%%%%%%%%%%%%%%%%%%%%%%%%%%%%%%%%%%%%%%%
\section{Methods}\label{sec:MET}
%%%%%%%%%%%%%%%%%%%%%%%%%%%%%%%%%%%%%%%%%%%%%%%%%%%%%%%%

We adapt the method \citep{Palumbo2020} proposed for the analysis of EHT measurements of the polarized emission around the supermassive black hole in M87 \citep{EHT2021_VII,EHT2021_VIII}. Here, we summarize the method and describe how this linear polarimetric decomposition can be used to estimate the underlying B-field structure of a spiral galaxy.

%%%%%%%%%%%%%%%%%%
%%%% FIGURE 1 %%%%
%%%%%%%%%%%%%%%%%%
\begin{figure}%[ht!]
\centering
\includegraphics[width=\columnwidth]{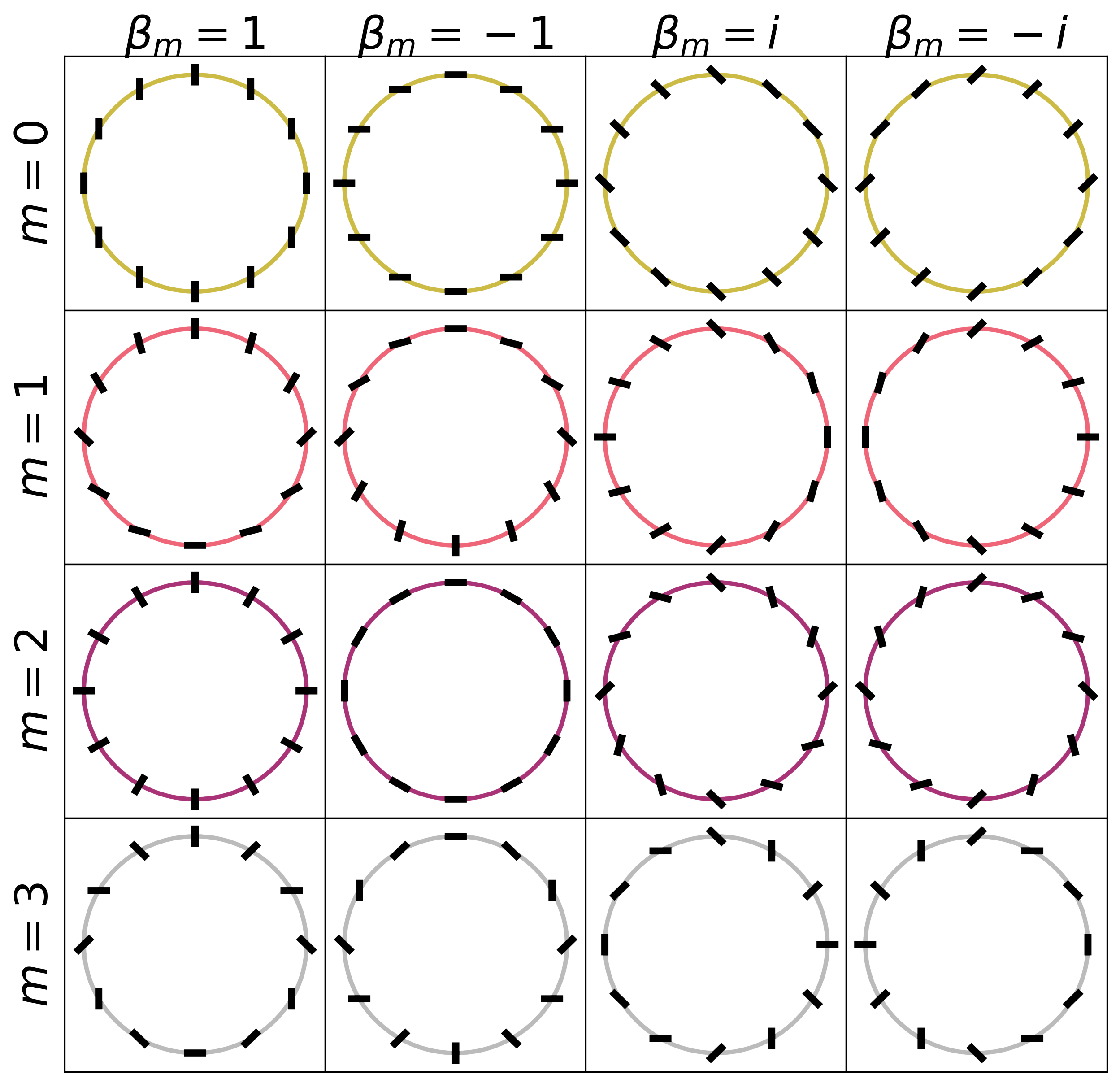}
\caption{Examples of the decomposition of the linear polarization fields in a ring. The morphology of the linear polarization field corresponding to $0 \leq m \leq 3$ modes with different values of the $\beta_m$ coefficient are presented.}
 \label{fig:fig1}
\end{figure}
%%%%%%%%%%%%%%

\subsection{Decomposition into azimuthal B-Field modes}\label{subsec:ModelDefinition}

We start by describing the linear polarization via the complex polarized intensity 

\begin{equation}\label{eq:complexP}
    P_{\rm B}(\rho, \phi) \equiv -Q(\rho, \phi) - iU(\rho, \phi),
\end{equation}
\noindent
where $Q$ and $U$ are the Stokes parameters of linear polarization, and $\rho$ and $\phi$ are radial and azimuthal coordinates, respectively. The sign convention in the definition of $P_{\rm B}$ represents our interest in the B-field orientation, which is rotated by $90^{\circ}$ from the electric vector position angle measured in radio and FIR polarimetric observations. %\textbf{Note that Eq. \ref{eq:complexP} provides the same information as the typical $Q-U$ diagrams used in studies of the interstellar polarization.} 
The measured polarization field is decomposed into azimuthal modes, $m$, with amplitudes of $\beta_m$ via the decomposition definition

\begin{equation}\label{1}
\beta_m = \frac{1}{I_{\rm ann}} \int_{\rho_{\rm min}}^{\rho_{\rm max}} \int_0^{2\pi} P_{\rm{B}}(\rho, \phi) e^{-im\phi} \rho d\phi d\rho,
\end{equation}
\noindent
where $I_{\rm{ann}}$ is the total Stokes I in the annulus range of $[\rho_{\rm{min}}, \rho_{\rm{max}}]$ defined as

\begin{equation}\label{eq:Iann}
I_{\rm{ann}} = \int_{\rho_{\rm min}}^{\rho_{\rm max}} \int_0^{2\pi} I(\rho, \phi) \rho d\phi d\rho.
\end{equation}

$\beta_m$ is a dimensionless complex number. Its absolute value, $|\beta_{\rm m}|$, corresponds to the amount of coherent power in the $m^\mathrm{th}$ mode, and the argument, $\angle \beta_{\rm m}$, corresponds to the average pointwise rotation of the B-field orientation within an annulus of radius $[\rho_{\rm{min}},\rho_{\rm{max}}]$. We define $\phi = 0^{\circ}$ as the B-field orientation in the north direction with positive values increasing along the counterclockwise direction (East of  North). This decomposition can be thought of as an azimuthally averaged Fourier transform of the complex polarization field per annulus, where the $\beta_m$ coefficients are Fourier coefficients corresponding to the internal Fourier modes. 

We provide a collection of examples of linear polarization fields in a ring corresponding to $0 \leq m \leq 3$ modes with different values of the $\beta_m$ coefficient in Figure \ref{fig:fig1} \citep[see also figure 1 of][]{Palumbo2020}. Figure \ref{fig:fig1} shows the B-field structure for the $0 \leq m \leq 3$ periodic modes with different values of the $\beta_m$ coefficient. Each polarization field component has an absolute amplitude of  $|\beta_{\rm{m}}| = 1$. The different configurations arise from the mode, $m$, and sign of $\beta_{\rm{m}}$. The B-field orientations along the ring are offset by an angle $\angle \beta_m$ given by half of the complex phase of $\beta_m$ $=(\Re{(\beta_m)} + \Im{(\beta_m}$)) within $[-90, 90]^{\circ}$

\begin{equation}\label{2}
\angle \beta_m = \frac{1}{2}\arctan{\left( \frac{\Im{(\beta_m)}}{\Re{(\beta_m)}} \right)}.
\end{equation}

The $m = 0$ mode corresponds to a constant B-field orientation, $m = 1$ mode corresponds to a half dipole field structure, and $m = 2$ corresponds to a radial and toroidal distribution in the real space and a spiral structure in the complex space. Note that the $m=2$ mode is analogous to the $E$ and $B$ mode decomposition commonly used in studies of CMB polarization \citep[e.g.,][]{Kamionkowski1997,SK1997,Zaldarriaga2001}, where the real part of $\beta_2$ is the $E$-mode and the imaginary part is the $B$-mode. We show the reconstruction of a non-trivial B-field orientation with a combination of $m=0$ and $m=2$ modes in Figure \ref{fig:fig2}. 

%%%%%%%%%%%%%%%%%
%%%% TABLE 1 %%%%
%%%%%%%%%%%%%%%%%
\begin{deluxetable*}{lcccccl}
\centering
\tablecaption{Galaxy Sample. \emph{Columns, from left to right:} (a) Galaxy name. (b) Galaxy distance in Mpc. (c) Physical scale in pc per arcsec. (d) Galaxy type. (e) Inclination of the galaxy in degrees. (f) Position angle of the long axis of the galaxy in the plane of the sky. (g) References for the distance, inclination, and tilt angles. 
\label{tab:GalaxySample} 
}
\tablecolumns{6}
\tablewidth{0pt}
\tablehead{\colhead{Galaxy} & 	\colhead{Distance$^{1}$}  & \colhead{Scale} & \colhead{Type$^{\star}$} & 
\colhead{Inclination (i)$^{2}$} &	\colhead{Tilt (PA)$^{2}$} &  \colhead{References} \\ 
 	&  \colhead{(Mpc)}	& \colhead{(pc/\arcsec)}	&
\colhead{($^{\circ}$)} & \colhead{($^{\circ}$)} & \colhead{($^{\circ}$)} \\
\colhead{(a)} & \colhead{(b)} & \colhead{(c)} & \colhead{(d)} & \colhead{(e)} & \colhead{(f)} & \colhead{(g)}} 
\startdata
M51 		&	$8.58$	&	$41.21$	&	Sa			&	$22.5\pm5$	&	$-7\pm3$		&
$^{1}$\citet{McQuinn2017}; $^{2}$\citet{Colombo2014}	\\
M83 		&	$4.66$	&	$22.38$	&	SAB(s)c		&	$25\pm5$		&	$226\pm5$	&
$^{1}$\citet{Tully2013}; $^{2}$\citet{Crosthwaite2002}	\\
NGC~3627 	&	$8.90$	&	$42.75$	&	SAB(s)b		&	$52\pm1$		&	$176\pm1$	&
$^{1}$\citet{Kennicutt2003}; $^{2}$\citet{Kuno2007} 		\\
NGC~4736 	&	$5.30$	&	$25.46$	&	SA(r)ab		&	$36\pm7$		&	$292\pm2$	&	
$^{1}$\citet{Kennicutt2003}; $^{2}$\citet{Dicaire2008} 		\\
NGC~6946 	&	$6.80$	&	$32.66$	&	Sc			&	$38.4\pm3.0$	&	$239\pm1$	&
$^{1}$\citet{Karachentsev2000}; $^{2}$\citet{Daigle2006} \\
\enddata
\tablenotetext{{\star}}{Galaxy type from NASA/IPAC Extragalactic Database (NED; \url{https://ned.ipac.caltech.edu/})}
\end{deluxetable*}

\subsection{Implementation of the algorithm}\label{subsec:ModelSteps}

We estimate the B-field orientation over a spiral galaxy as follows.

\begin{figure}
\includegraphics[width=\columnwidth]{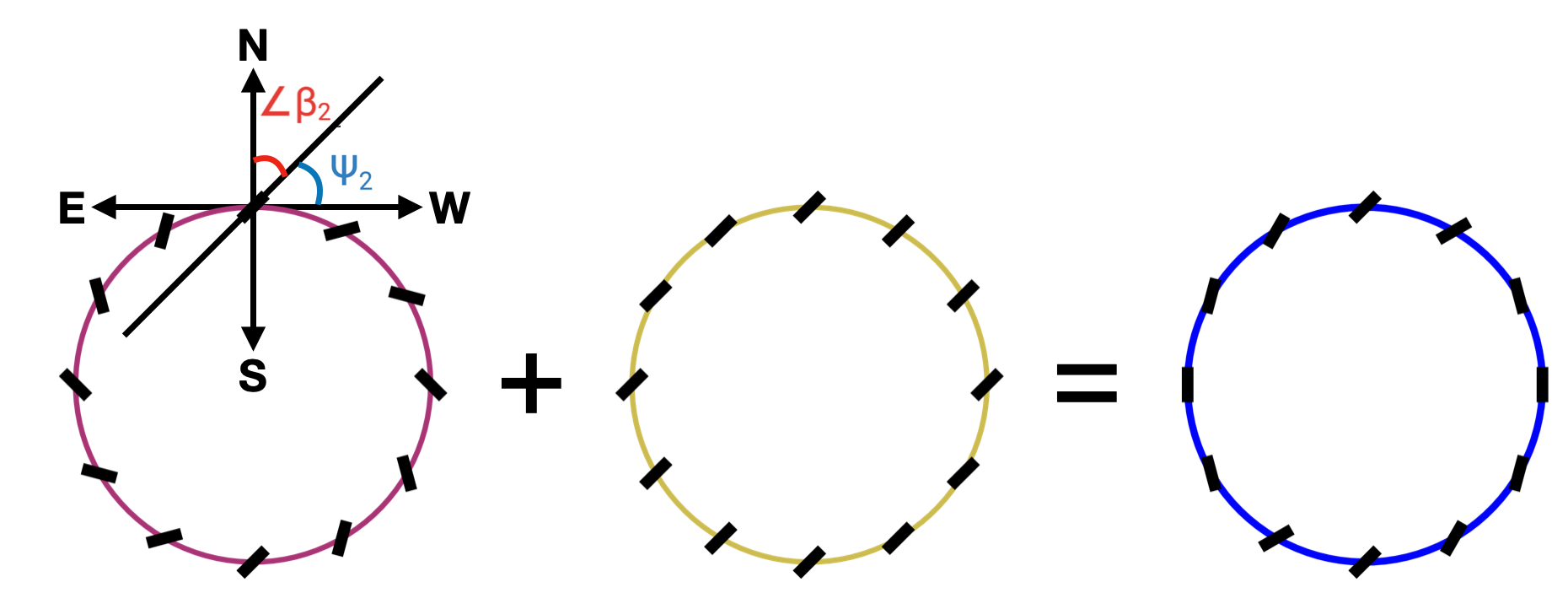}
\caption{Example of the B-field decomposition of a galaxy. The composition of the $m = 2$ mode with ${\beta_2} = -i$ and the $m = 0$ mode with ${\beta_0} = -i$ is shown. Additionally, the relationship between the averaged pointwise rotation of the B-field orientation, $\angle{\beta_2}$, and the magnetic pitch angle, $\Psi_2$, is illustrated at the top of the $m = 2$ mode on the left.
 \label{fig:fig2}}
\end{figure}

\begin{enumerate}
    \item We construct a two-dimensional map of azimuthal angles such that $\phi=0^{\circ}$ corresponds to North (up), and $\phi$ increases in the counterclockwise direction (East from North).
    
    \item We define a set of radial masks centered at the peak of the galaxy's central emission at a given wavelength. We use the same central point for both FIR and radio wavelengths. The first step is to define the grid of radial distances, i.e., the projected distance of each pixel from the galaxy's center in the $x, y$ plane (with the line of sight along $z$). For a galaxy perfectly face-on, the radial mask is a simple circular annulus centered at the peak of the galaxy. For inclined galaxies, we rotate the radial mask by the galaxy's inclination, $i$, and tilt, $\theta$, angles. We calculate $r' = R_x[i]R_z[\theta]r$, where $r$ is the original radius, $r'$ is the new radial distance, and $R_x[i]$, $R_z[\theta]$ are the rotation matrices for the inclination and tilt, respectively. We can thus define the projected annulus for any given inner and outer radius.  
    
    \item We compute $P_{\rm B}(\rho, \phi) \equiv -Q(\rho, \phi) - iU(\rho, \phi)$, the complex-valued polarized intensity (Equation \ref{eq:complexP}), where $Q$ and $U$ are the measured Stokes parameters at the galactrocentric radius of $\rho = \sqrt{x^{2} + y^{2}}$ and azimuthal angle $\phi$. 
    
    \item Using Equation \ref{1}, we calculate the amplitude $|\beta_m|$ and angle $\angle{\beta_m}$ for every annulus.
 
    \item For $m = 2$ only, we take the product of the basis function $e^{im\phi}$ and the complex-valued polarization field $P(\rho, \phi) = Q(\rho, \phi) + iU(\rho, \phi)$. Note that this definition of the polarized intensity is the opposite sign to Equation \ref{eq:complexP}. We define this for $m=2$ to facilitate comparison with other measurements of galaxy magnetic pitch angles: this quantity represents the tangent to the local circumference at a given distance from the galaxy’s center, which is equivalent to the pitch angle of the B-field, $\Psi_2$. The angles $\angle \beta_{2}$ and $\Psi_{2}$ are thus complementary. Figure \ref{fig:fig2} illustrates the definition of each.
\end{enumerate}

We estimate the uncertainty on our decomposition parameters via a Monte Carlo technique. We generate $5000$ realizations of the Stokes $I$, $Q$, and $U$ fields by randomly sampling each pixel from a Gaussian distribution centered on the measured value, and with a standard deviation equal to the measurement uncertainty. From each realization we compute $|\beta_m|$, $\angle{\beta_m}$, and $\Psi_{2}$ for each annulus defined by radial range $[\rho_{\rm{min}}, \rho_{\rm{max}}]$. We compute the mean and standard deviation of each quantity over the $5000$ samples. This method is implemented in \textsc{python}, and the code is available in Zenodo\footnote{The code used in this work is registered with DOI \url{https://zenodo.org/record/8011542}}. 

%%%%%%%%%%%%%%%%%%
%%%% FIGURE 2 %%%%
%%%%%%%%%%%%%%%%%%
%%%%%%%%%%%%%%

%%%%%%%%%%%%%%%%%%%%%%%%%%%%%%%%%%%%%%%%%%%%%%%%%%%%%%%%
\section{Application}\label{sec:APPLICATION}
%%%%%%%%%%%%%%%%%%%%%%%%%%%%%%%%%%%%%%%%%%%%%%%%%%%%%%%%

We describe the data used in this work and show the results of this method to quantify the B-field structure of spiral galaxies. 

\subsection{Archival data}\label{subsec:ArchivalData}

We apply the method presented in Section \ref{sec:MET} to a sample of five spiral galaxies. These spiral galaxies are the only publicly available objects with combined FIR and radio polarimetric observations. The application to both wavelengths allows us to characterize the B-field morphology in two different phases of the ISM. Table \ref{tab:GalaxySample} lists the properties of the galaxy sample used in this work. 

The FIR data were taken from the Survey of ExtragALactic magnetiSm with SOFIA (SALSA\footnote{Data from SALSA can be found at \url{http://galmagfields.com/}}) published by \citet[][SALSA IV]{SALSAIV}. All FIR polarimetric observations were performed using SOFIA/HAWC+ at $154$ \um~with a beam size (FWHM) of $13.6$\arcsec~and a pixel scale of $6.90$\arcsec~(i.e., Nyquist sampling). For a detailed description of the data reduction see \citet[][SALSA III]{SALSAIII} and for an analysis of the polarization fraction see \citet[][SALSA IV]{SALSAIV} and B-field orientation see \citet[][SALSA V]{SALSAV}.

The radio polarimetric observations were obtained with the Very Large Array (VLA) and the Effelsberg 100-m radio telescope at $6$ cm with a typical angular resolution of $8$\arcsec. For a detailed description of the data reduction of each galaxy we defer to \citet[][M51]{Fletcher2011}, \citet[][M83]{Frick2016}, \citet[][NGC~3627]{Soida2001}, \citet[][NGC~4736]{Chyzy2008}, and \citet[][NGC6946]{Beck1991,Beck2007}. The $6$ cm polarimetric observations were selected because they are the common radio wavelength of all galaxies and have higher signal-to-noise ratios than the $3$ cm. At longer radio wavelengths ($18$, $20$ cm), the observations can be strongly affected by Faraday rotation \citep{Beck2019}. For all radio observations, the Stokes $I$, $Q$, and $U$ were convolved with a 2D Gaussian kernel to match the beam size of the HAWC+ observations. Then each galaxy was reprojected to match the footprint and pixelization of the HAWC+ observations. The smoothed and reprojected radio polarimetric observations are publicly available on the SALSA website. Figure \ref{fig:fig3} shows the measured B-field orientation at $154$ \um~and $6$ cm for each of the spiral galaxies used for our study. For visualization purposes, we display one B-field orientation per beam and only polarization measurements with $PI/\sigma_{PI} \ge 3$, where $PI$ is the polarized intensity and $\sigma_{PI}$ is the associated uncertainty. The polarization measurements used for the analysis are described in Section \ref{subsec:results}.

%%%%%%%%%%%%%%%%%%%%%%%%%%%%%%%%%%%%%%%%%%%%%%%%%%%%%%%%%
\subsection{Results of the B-field orientation decomposition}\label{subsec:results}
%%%%%%%%%%%%%%%%%%%%%%%%%%%%%%%%%%%%%%%%%%%%%%%%%%%%%%%%

%%%%%%%%%%%%%%%%%%
%%%% FIGURE 3 %%%%
%%%%%%%%%%%%%%%%%%
\begin{figure*}
\centering
\includegraphics[width=0.85\textwidth]{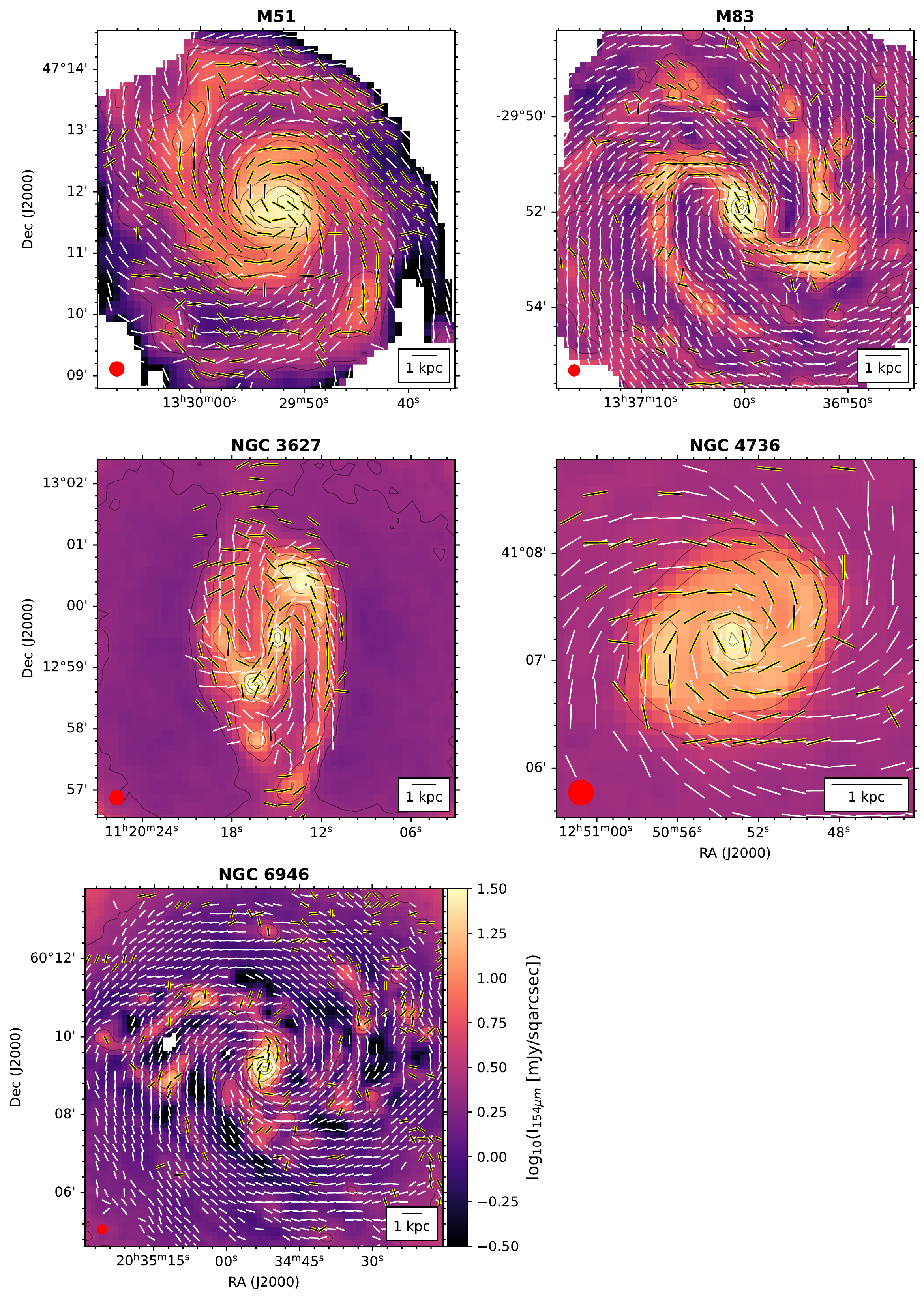}
\caption{B-field orientation of our sample of spiral galaxies. Polarization measurements are shown with constant
length to illustrate the inferred B-field orientations at $154$ \um~(black lines) and $6$ cm (white lines) overlaid on the $154$~\um~total intensity (color scale). All figures share the same colorscale as shown in the colorbar of NGC~6946. The contours are in fractions of $0.2$, $0.4$, $0.6$, $0.8$, $0.9$, $0.98$ of the peak intensity. Polarization measurements per beam (red circle) with $PI/\sigma_{PI} \ge 3.0$ are shown, where $PI$ and $\sigma_{PI}$ are the polarized intensity and its uncertainty, respectively. A legend with the spatial scale of $1$ kpc is shown in each panel.
 \label{fig:fig3}}
\end{figure*}
%%%%%%%%%%%%%%

\begin{figure*}%[ht!]
\includegraphics[width=\textwidth]{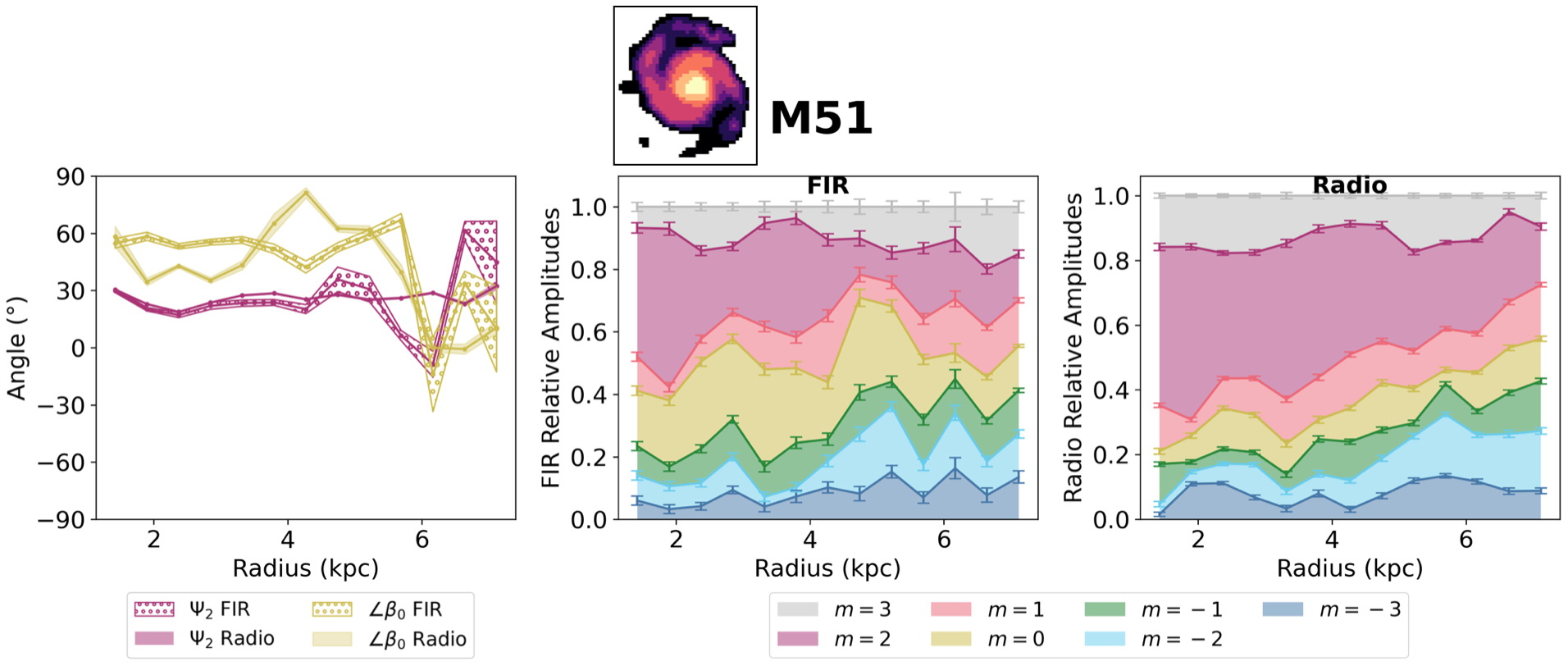}\
\includegraphics[width=\textwidth]{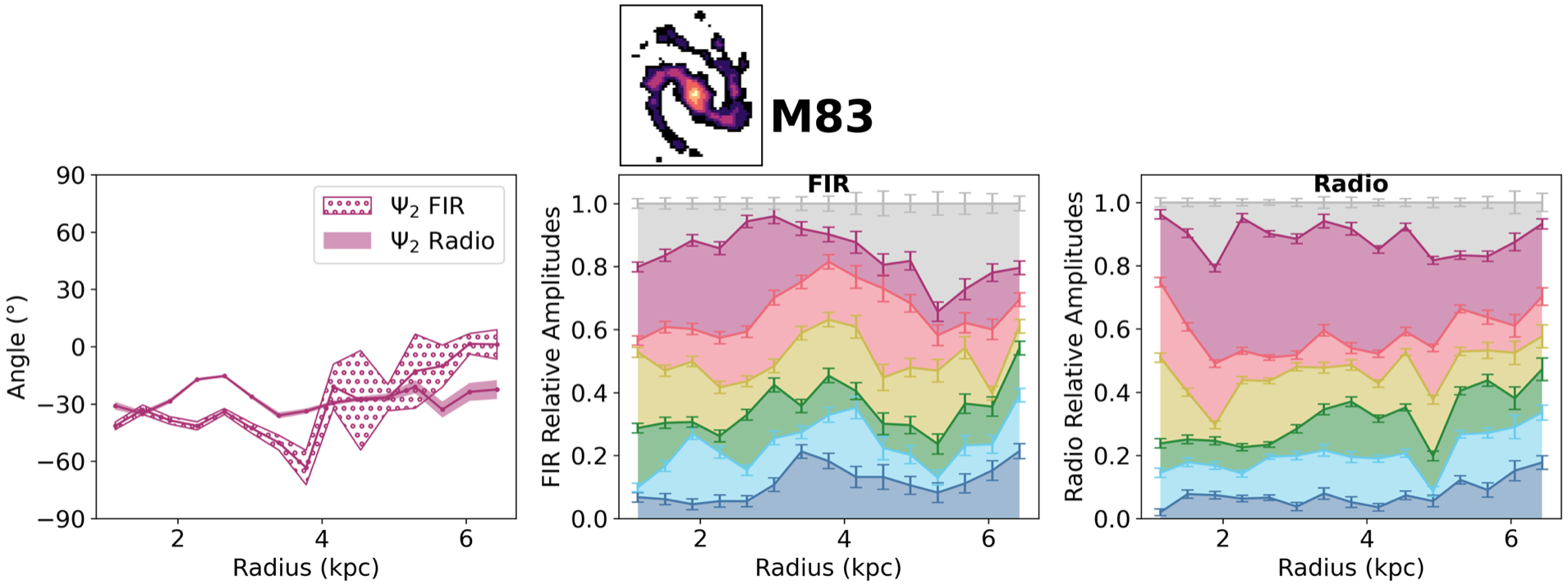}\
\caption{Linear decomposition results of the FIR and radio B-fields for M51 and M83. The titles of each galaxy show the intensity map at FIR wavelengths to guide the reader with the morphological structure of the galaxy. We show the FIR and radio magnetic pitch angles, $\Psi_2$, as a function of the galactocentric radius with the middle solid line corresponding to the mean values and the width of the line corresponding to $\pm 1\sigma$ (Left). Note that for M51, we have included $\angle \beta_0$ which is the angle associated with the $m = 0$ mode. The center panel shows the relative amplitude of each mode as a function of distance from the galactic center at FIR wavelengths with the width of each band corresponding to the value of the mode's relative amplitude. Additionally, on the top of each band for each mode are error bars, showing $\pm 1\sigma$. The right panel shows the same but at radio wavelengths. 
\label{fig:fig4}}
\end{figure*}

\begin{figure*}[ht!]
\includegraphics[width=\textwidth]{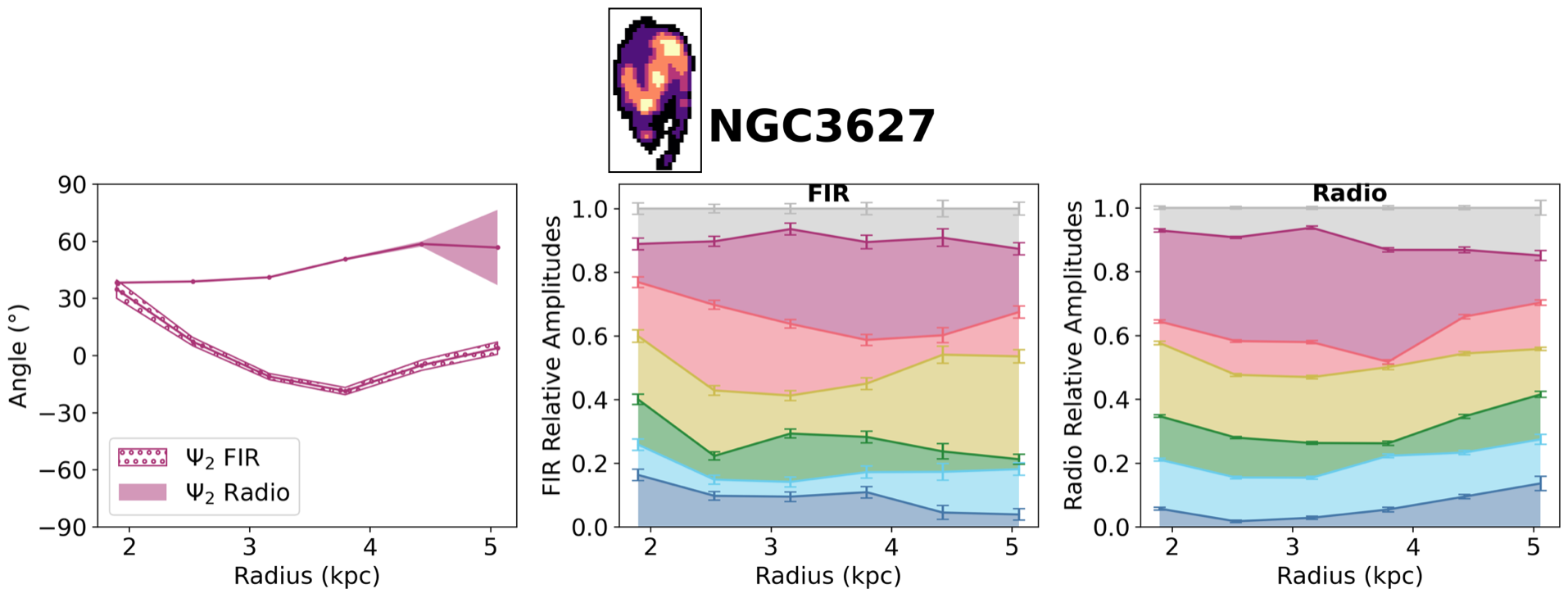}\
\includegraphics[width=\textwidth]{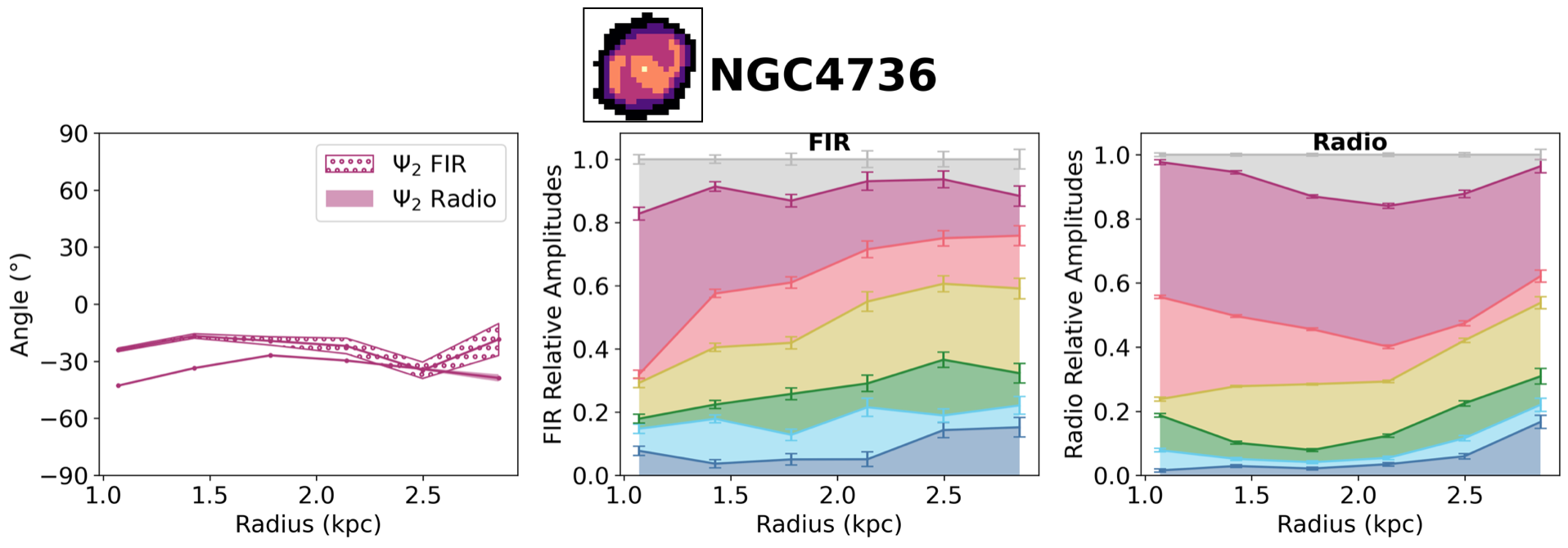}\
\includegraphics[width=\textwidth]{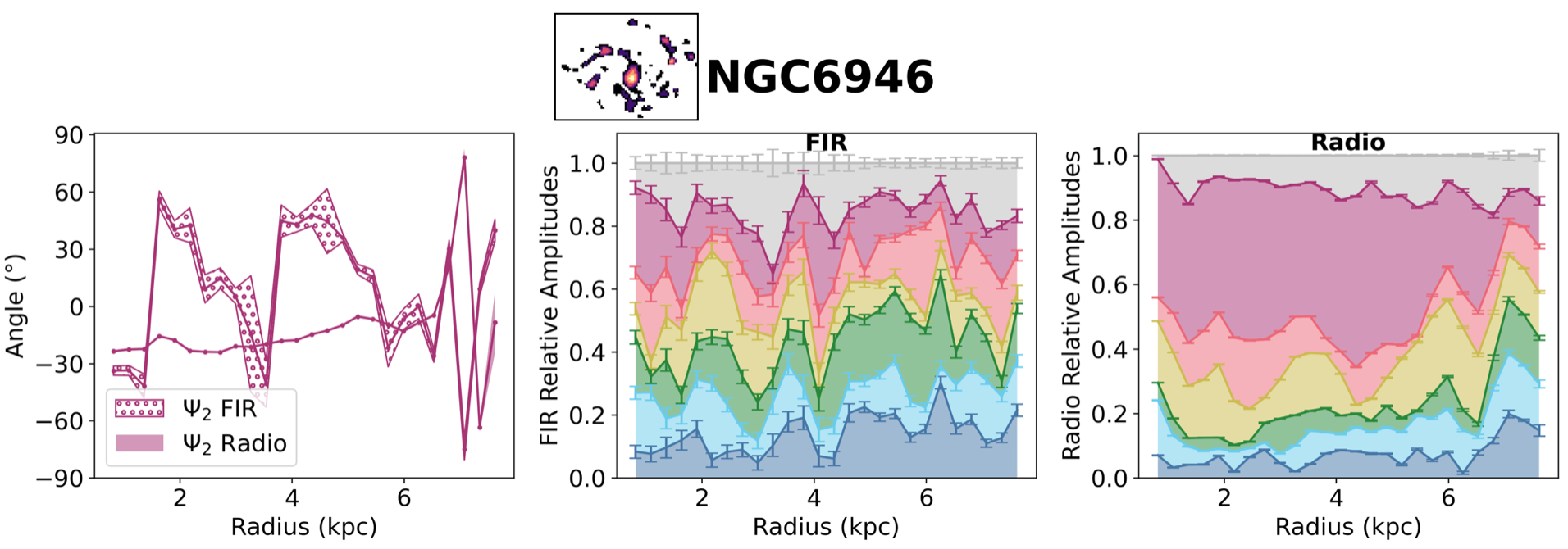}
\caption{As in Figure \ref{fig:fig4} with NGC~3627, NGC~4736, and NGC~6946.
 \label{fig:fig5}}
\end{figure*}

We apply the steps presented in Section \ref{subsec:ModelSteps} to the five galaxies shown in Figure \ref{fig:fig3}. For the analysis of the linear polarimetric decomposition, we select data with $I/\sigma_{I} \ge 10$, where $I$ and $\sigma_{I}$ are the Stokes $I$ and its uncertainty, respectively. Table \ref{tab:GalaxySample} shows the galaxy’s inclination, $i$, and tilt, $\theta$, angles used to define the projected annuli. We calculate the radial profiles selecting the width of each annulus to be equal to the beam size of the HAWC+ observations, i.e., $13\farcs6$ ($2$ pixels). The core ($2$ beams = $27\farcs2$ = $4$ pixels) of each galaxy is masked because of the limited number of independent measurements in that innermost region. Each decomposition is centered at the location of the peak total intensity of the radio emission. All of these galaxies have an unresolved core at radio wavelengths, while the FIR emission shows an extended core (e.g., M83) or a dearth of central emission (e.g., M51). We test the robustness of the central coordinate selection by varying the central coordinates by $\pm1$ pixel in all directions.  Specifically, we moved the central coordinates by $\pm1$ pixel and calculated the standard deviation of all measurements to be $<8^{\circ}$ at FIR wavelengths and $<4^{\circ}$ at radio wavelengths in the final pitch angles, $\Psi_{2}$, across the entire disk. We show the results (Figures \ref{fig:fig4} and \ref{fig:fig5}) out to the largest radius where we can measure an uncertainty on $\Psi_{2}$ $\le30^{\circ}$. Figures \ref{fig:fig4} and \ref{fig:fig5}  display the intensity map at FIR wavelengths to show the morphological structure of the galaxy. These figures show the magnetic pitch angles and relative amplitudes at radio and FIR wavelengths as a function of the galactocentric radius. We describe the amplitudes and magnetic pitch angles in the following sections.

%%%%%%%%%%%%%%%%%%%%%%%%%
%%%% FIGURES 4 and 5 %%%%
%%%%%%%%%%%%%%%%%%%%%%%%%

%%%%%%%%%%%%%%

\subsubsection{Amplitudes}\label{subsubsec:amplitudes}

We compute $\beta_{m}$ for the $-3 \le m \le 3$ modes and show their relative amplitudes as a function of the radii (Figures \ref{fig:fig4} and \ref{fig:fig5}). We estimate the statistical significance of B-field modes in the $m=[-17,17]$ range (Appendix \ref{App:A1}). We find that the $m=[-3,3]$ range is the most representative set of B-field modes in our galaxy sample. This result is mainly because the complexity of the B-field morphology described by higher modes is at smaller scales than the resolution of the observations. The fractional amplitude per annulus, per galaxy is 

\begin{equation}
    \tilde{|\beta_{m}|} = \frac{|\beta_{m}|}{\sum_{m=-3}^{m=3}|\beta_{m}|},
\end{equation}
where the uncertainties are estimated using the Monte Carlo approach described in Section \ref{subsec:ModelSteps}. 

\begin{figure}[ht!]
\includegraphics[width=\columnwidth]{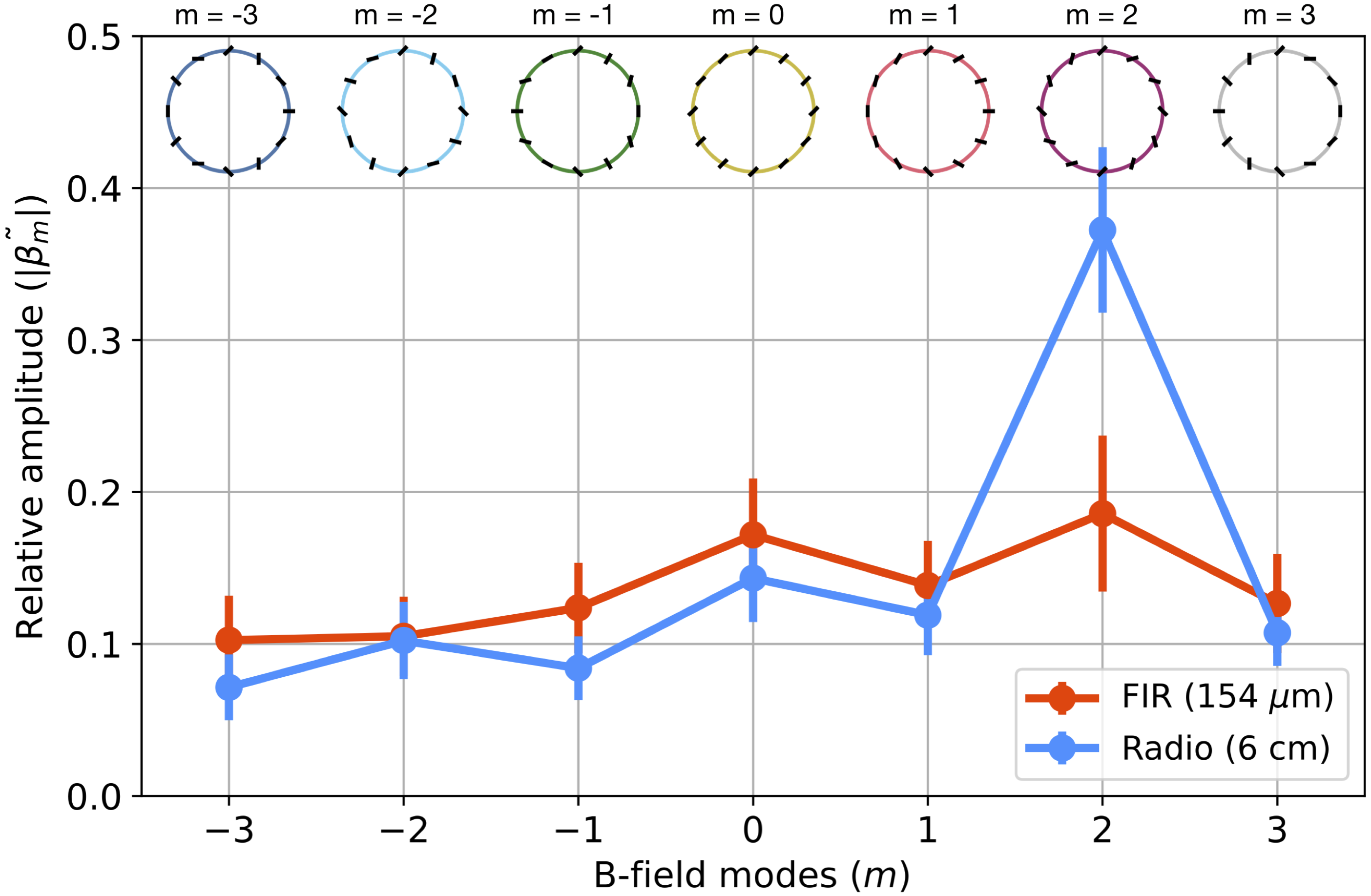}
\caption{Mean relative amplitudes of the B-field modes of a composited spiral galaxy. FIR (red) and radio (blue) relative amplitudes for modes $-3\le m \le 3$ are shown. The B-field pattern associated with each mode is shown at the top.
 \label{fig:fig6}}
\end{figure}

Figures \ref{fig:fig4} and \ref{fig:fig5} show that the most common mode is $m=2$ at radio wavelengths. The $m=2$ is also common at FIR wavelength, except for NGC~6946. For NGC~6946 all modes across the full disk are equally important. To quantify these trends, we can further average over all annuli and all galaxies and estimate the mean relative mode amplitude of our spiral galaxies (Figure \ref{fig:fig6} and Table \ref{tab:beta_m}). 

At radio ($6$ cm) wavelengths, we find that $m=2$ is the most dominant mode for our spiral galaxies. We estimate that the dominant $m=2$ mode at $6$ cm has a mean relative contribution of $\tilde{|\beta_{2}|} = 0.37\pm0.05$, averaged over annuli. The following modes are $m=0$, $m=1$, and $m=3$ with similar, within the uncertainties, relative contributions to the radio B-field of $\tilde{|\beta_{0}|} = 0.14\pm0.03$, $ \tilde{|\beta_{1}|} = 0.12\pm0.03$, and $ \tilde{|\beta_{3}|}= 0.11\pm0.02$, respectively. All negative modes have relative amplitudes  $<0.1$ in the radio, but together they account for $\tilde{|\beta_{m<0}|} \sim 0.25$ of the total average mode amplitude.

At FIR ($154~\mu$m) wavelengths, $m=2$ and $m=0$ have similar, within the uncertainties, relative contributions of $\tilde{|\beta_{2}|} = 0.19\pm0.05$ and $\tilde{|\beta_{0}|} = 0.17\pm0.04$ averaged over the full disk. The modes $m=1$ and $m=3$ have the same relative amplitude of $\sim0.13$. The negative modes have relative amplitudes in the range of $0.10-0.12$, and they sum to $|\tilde{\beta}_{m<0}| \sim 0.33$ of the total average FIR polarization mode amplitude. The FIR polarization data thus show significantly weaker mode preference compared to the radio polarization data in general. This is evident in the individual galaxy mode decompositions (Figures \ref{fig:fig4} and \ref{fig:fig5}), as well as the mean relative mode contribution for each data set (Figure \ref{fig:fig6}).  

NGC~6946 displays the most extreme difference between the mode decomposition in the radio and the FIR data (Figure \ref{fig:fig5}, last row). In the radio, the $m=2$ mode is strongly dominant, with $\tilde{|\beta_{2}|}= 0.43\pm0.06$. By contrast, all modes contribute roughly equally to the FIR polarization ($\tilde{|\beta_{m}|} = [0.11-0.17]$). This is perhaps not surprising because of the galaxies in our sample, NGC~6946 has the most irregular B-field morphology in the FIR. We recalculate the mean relative mode amplitudes for the four galaxies excluding NGC~6946, and we find that the Table \ref{tab:beta_m} values do not change within the stated uncertainties. 

%%%%%%%%%%%%%%%%%%
%%%% FIGURE 6 %%%%
%%%%%%%%%%%%%%%%%%
%%%%%%%%%%%%%%

%%%%%%%%%%%%%
%%%% TABLE 2 %%%%
%%%%%%%%%%%%%
\begin{deluxetable}{lccccc}
\tablecaption{Mean relative amplitudes of a spiral galaxy. \emph{Columns, from left to right:} 
(a) Mode.
(b) FIR relative amplitude. 
(c) Radio relative amplitude. The errors represent the standard deviation of the pitch angle profile, not the uncertainties of the average value.
\label{tab:beta_m} 
}
\tablecolumns{9}
\tablewidth{0pt}
\tablehead{\colhead{Mode} & 	 \colhead{$\langle |\tilde{\beta_{m}^{\rm{FIR}}} |\rangle$} &  \colhead{$\langle |\tilde{\beta_{m}^{\rm{Radio}}} |\rangle$} 
} 
\startdata
3 	&	$0.13\pm0.03$   &   $0.11\pm0.02$	\\
2 	&	$0.19\pm0.05$   &   $0.37\pm0.05$	\\
1 	&	$0.14\pm0.03$   &   $0.12\pm0.03$		\\
0 	&	$0.17\pm0.04$   &   $0.14\pm0.03$ 		\\
-1 	&	$0.12\pm0.03$   &   $0.08\pm0.02$ \\
-2 	&	$0.11\pm0.03$   &   $0.10\pm0.03$ \\
-3 	&	$0.10\pm0.03$   &   $0.07\pm0.02$ 
\enddata
\end{deluxetable}

\subsubsection{Magnetic pitch angles}\label{subsubsec:pitchangles}

The magnetic pitch angle, $\Psi_{2}$, is the complementary angle to the $\angle \beta_{2}$ angle estimated by the B-field mode decomposition method (Figure \ref{fig:fig2}). We show radial profiles of the pitch angles for each galaxy in Figures \ref{fig:fig4} and \ref{fig:fig5}. We also estimate the mean pitch angles, $\langle \Psi_{2} \rangle$, per galaxy at FIR and radio wavelengths (Table \ref{tab:pB}). Figure \ref{fig:fig4} also includes $\angle \beta_{0}$ (the azimuthal angle associated with the $m=0$ mode) for M51 because it has a high contribution in the B-field modes of this galaxy.

We estimate that the mean pitch angle at FIR wavelengths is smaller than at radio wavelengths (Table \ref{tab:pB}), i.e., $\tilde{\langle |\Psi_{2}^{\rm{FIR}}|\rangle} < \tilde{\langle |\Psi_{2}^{\rm{Radio}}|\rangle}$. This result indicates that radio spiral B-fields are more open than FIR spiral B-fields in our sample. In addition, the FIR wavelengths have pitch angles with a larger angular dispersion, $\pm24^{\circ}$, than at radio wavelengths, $\pm8^{\circ}$. This result shows that radio spiral B-fields are more ordered than FIR spiral B-fields.

At radio ($6$ cm) wavelengths, we estimate that $\Psi_{2}^{\rm{Radio}}$ increases as the galaxy radius increases for M51, NGC~3627, while $\Psi_{2}^{\rm{Radio}}$ decreases for M83, NGC~4726, and NGC~6946. This result is consistent with the literature \citep{Beck2019, SALSAV} and implies that $\Psi_{2}^{\rm{Radio}}$ opens up toward the outskirts of the disk for positive magnetic pitch angles and is wrapped tigher for negative magnetic pitch angles. It is interesting to note the drastic change of the pitch angle from $-30^{\circ}$ to $-10^{\circ}$ at a projected radial distance of $\sim2.5$ kpc in M83 (Figure \ref{fig:fig4}). At $\sim2.5$ kpc, the pitch angle varies in the interface between the bar region and the spiral arms as determined from the velocity fields of gas tracers (i.e., HII, CO) and stellar dynamics \citep{Kenney1991}. 

At FIR ($154$ \um) wavelengths, $\Psi_{2}^{\rm{FIR}}$ shows many variations as a function of the galactrocentric radius, although in all cases except NGC~6946 there is a large-scale spiral ordered B-field evident in the FIR polarization. We find that at certain radii (e.g. $2.5-3.5$ kpc) in NGC~6946 the FIR B-field has similar pitch angles to those measured at radio wavelengths (Fig. \ref{fig:fig5}). However, the angular dispersion of NGC~6946 is large, $\pm41^{\circ}$, across the disk when compared to the radio B-fields, $\pm5^{\circ}$ (Table \ref{tab:pB}). 

%%%%%%%%%%%%%
%%%% TABLE 3 %%%%
%%%%%%%%%%%%%
\begin{deluxetable*}{lcccccl}
\tablecaption{Mean magnetic pitch angle per galaxy and wavelength from this work vs. literature. \emph{Columns, from left to right:} 
(a) Galaxy name.
(b) FIR magnetic pitch angle. 
(c) Radio magnetic pitch angle.
(d) Range of radio magnetic pitch angle.
(e) Radio magnetic pitch angles of the ordered B-fields from the literature. 
(f) Range of radio magnetic pitch angles of the ordered B-fields from the literature.
(g) References of (d, e). In this table, the errors represent the standard deviation of the pitch angle profile, not the uncertainties of the average value.
\label{tab:pB} 
}
\tablecolumns{9}
\tablewidth{0pt}
\tablehead{\colhead{Galaxy} & 	 
\colhead{$\langle \Psi_{2}^{\rm{FIR}} \rangle$} &  
\colhead{$\langle \Psi_{2}^{\rm{Radio}} \rangle$}  & 
\colhead{$\Psi_{2}^{\rm{Radio}}$}  & 
\colhead{$\langle p_{\rm{o}}^{\rm{Radio}} \rangle$} & 
\colhead{ $p_{\rm{o}}^{\rm{Radio}}$} &
\colhead{References}  \\ 
  & \colhead{($^{\circ}$)} & \colhead{($^{\circ}$)} & \colhead{($^{\circ}$)}					\\
\colhead{(a)} & \colhead{(b)} & \colhead{(c)} & \colhead{(d)} & \colhead{(e)} & \colhead{(f)} & \colhead{(g)} } 
\startdata
M51 		&	$25\pm17$   &   $28\pm5$	& $[16-34]$ & $22\pm2$ & $[19-27]$ & \citet{Fletcher2011}\\
M83 		&	$-41\pm17$   &   $-29\pm8$	& $-[12-36]$ & $-30\pm3$ & $-[23-35]$ & \citet{Beck2019}\\
NGC~3627 	&	$4\pm17$   &   $49\pm8$		& $[41-51]$ & $37\pm8$ & $[16-68]$ & \citet{Soida2001}\\
NGC~4736 	&	$-21\pm6$   &   $-30\pm3$ 	& $-[24-32]$ & $-35\pm5$ & -	& \citet{Chyzy2008} \\
NGC~6946 	&	$6\pm41$   &   $-17\pm5$    & $-[8-25]$ & $-27\pm2$ & $[30-32]$ & \citet{Ehle1993,Beck2019}\\
$\langle\tilde{|\Psi_{2}|\rangle}$ & $21\pm24$  &    $29\pm8$ & - & $30\pm7$ & -
\enddata
\end{deluxetable*}

%%%%%%%%%%%%%%%%%%%%%%%%%%%%%%%%%%%%%%%%%%%%%%%%%%%%%%%%
\section{Discussion}\label{sec:DIS}
%%%%%%%%%%%%%%%%%%%%%%%%%%%%%%%%%%%%%%%%%%%%%%%%%%%%%%%%

%%%%%%%%%%%%%%%%%%%%%%%%%%%%%%%%%%%%%%%%%%%%%%%%%%%%%%%%
\subsection{Galactic dynamo}\label{subsec:GalDyn}
%%%%%%%%%%%%%%%%%%%%%%%%%%%%%%%%%%%%%%%%%%%%%%%%%%%%%%%%

We quantified the large-scale ordered B-fields in the disk of galaxies as a linear combination of axisymmetric modes. We found that B-fields with $m=2$ (spiral pattern) modes dominate at radio wavelengths, while $m=2$ and $m=0$ (constant) modes have similar contributions at FIR wavelengths. The FIR data show a larger relative contribution from higher modes than the radio wavelengths. We discuss these measurements in the context of galactic dynamo theory.

Turbulent dynamo theory explains the measured B-fields as a combination of fluctuation (or small-scale) dynamos and mean-field (or large-scale) dynamos \citep{Subramanian1998,BS2005,ss21}. In this picture, the large-scale B-fields are generated by differential rotation of the galaxy disk and turbulent helical motions. The turbulent B-fields are generated by turbulent gas motions at scales $\lesssim 50-100$ pc scales of energy injection by supernova explosions and stellar feedback \citep{Ruzmaikin1988,BS2005,Haverkorn2008}. Present-day FIR and radio polarimetric observations \citep[][]{Beck2019,SALSAIV} with spatial resolutions of $\ge100$ pc cannot resolve the turbulent B-fields in galaxies. The measured B-fields are dominated by the large-scale B-fields, although angular fluctuations of polarimetric properties across the disk can be estimated and compared to expectations for sub-beam-scale physics like star formation, shear, and shocks.

The total B-field can be described as the sum of a regular (or coherent) component and a random component \citep[e.g.,][]{Haverkorn:2015,Beck2019}. A well-defined B-field direction within the beam size of the observations is described as a regular B-field. The random B-field component may have spatial reversals within the beam of the observations, which can be isotropic or anisotropic. The directions of the isotropic random B-fields have the same dispersion in all spatial dimensions. An anisotropic random B-field has a well-defined average orientation in addition to sub-beam-scale reversals. Observationally, the combination of anisotropic and regular B-fields is known as ordered B-fields. Polarized radio synchrotron emission traces the ordered (regular and anisotropic random) B-fields in the plane of the sky, which depends on the strength and geometry of the B-fields, and the cosmic ray electron density. Regular B-fields can only be traced using Faraday rotation measures, which are sensitive to the direction of the B-field along the LOS. Polarized FIR dust emission is sensitive to the density-weighted line-of-sight average of the plane-of-sky B-field orientation, in addition to dust properties like column density and temperature. In this work, we have measured and quantified the large-scale ordered B-fields in spiral galaxies traced by both FIR and radio polarimetry.

As mentioned in Section \ref{sec:INT}, linear models of the mean-field dynamo decomposed the regular B-fields in Fourier series in the azimuthal angle. For the galaxies and wavelengths in our sample, this approach finds that the disk of M51 is dominated by $m_{\rm{d}}=0$ and  $m_{\rm{d}}=2$ with a relative amplitude of $0.72\pm0.06$ \citep{Fletcher2011}. The halo is dominated by $m_{\rm{d}} =1$ and also has $m_{\rm{d}}=2$ with a relative amplitude of $0.30\pm0.09$. M83 is dominated by $m_{\rm{d}}=1$ and has $m_{\rm{d}}=0$ with a relative amplitude of $0.43\pm0.3$ \citep{Beck2019}. NGC~6946 has similar contributions of $m_{\rm{d}}=0$ and $m_{\rm{d}}=2$ \citep{Ehle1993,Rohde1999}. The rotation measures of NGC~3627 did not show distinguishable patterns, which was attributed to Faraday rotation from an extended hot, low-density ionized magnetized halo \citep{Soida2001}. 

Table \ref{tab:pB} shows a comparison of the magnetic pitch angles between our measurements, $\Psi_{2}$, and the literature, $p_{\rm{o}}$. We show the mean and ranges of the ordered B-field pitch angles estimated using the B-field orientations from radio polarimetric observations obtained from the literature. The range of $p_{\rm{o}}$ in some of the galaxies shows the minimum and maximum values of the literature magnetic pitch angles within the galactocentric radii used in our analysis. We see that $p_{\rm{o}}$ is similar to our measured $\langle \Psi_{2}^{\rm{Radio}} \rangle$. They are not equal mainly because $p_{\rm{o}}$ is affected by several B-field modes. %Despite that the LOS associated with the FIR measurements is not
Even though the high-SNR FIR measurements are not well-sampled in the inter-arm regions (Figure \ref{fig:fig3}), the mean magnetic pitch angles across the disk at radio wavelengths are similar using both methods, $\langle \tilde{\Psi_{2}^{\rm{Radio}}} \rangle = 29\pm8^{\circ}$ vs. $\langle \tilde{p_{\rm{o}}} \rangle = 30\pm7^{\circ}$. This result implies that the mean radio B-fields across the disk may be dominated by the large-scale ordered B-field in the inter-arms, where the arms may be larger affected by the contributions from star formation activity. 

Galactic dynamo models provide the azimuthal wave number, $m_{\rm{d}}$, and their associated pitch angles for the disk B-field and the helical B-field. Because the large-scale regular B-field is modeled as a linear combination of logarithmic spiral modes, all dynamo modes are equal to our $m=2$ (spiral B-field). Although $m=2$ is dominant, our method shows that the measured B-field is a combination of several B-field patterns (Figure \ref{fig:fig6}). These modes (and the combination of them) can be interpreted as non-axisymmetric ordered B-fields showing deviations from the large-scale spiral ordered B-fields, perhaps due to particular physics (e.g., star-forming regions, shearing, compression, and/or shocks) across the disk. We note that the rotation measure distribution across a galaxy can also be used to measure pitch angles and B-field direction, providing complementary information \citep{Beck2019}.

%%%%%%%%%%%%%%%%%%%%%%%%%%%%%%%%%%%%%%%%%%%%%%%%%%%%%%%%
\subsection{Comparison with geometrical models}\label{subsec:Comp} 

The B-field morphologies in galaxies have also been quantified using pure geometrical models. We describe these methods and discuss their advantages and caveats.

\textit{Axisymmetric B-fields}: This approach estimates the pixel-by-pixel pitch angles across the galaxy disk. The measured B-field orientations are reprojected and tilted to obtain a face-on view of the galaxy, and an azimuthal template is subtracted from the data. The radial pitch angles are estimated as the mean of the pitch angles at a given annulus. This method was developed by \citet{SALSAI} and applied to the same M51 observations presented here. The advantages of this method are that a) the pitch angles on a pixel-to-pixel basis can be estimated, and b) the pitch angles are estimated without prior assumptions about the morphology of the B-field pattern. The pixel-by-pixel map can be used to estimate the means of the magnetic pitch angles from specific areas of the disk, like spiral arms or inter-arm regions, by applying user-defined masks \citep{SALSAI}. The disadvantage is that the angular offsets between the measured B-field orientations and the azimuthal profiles are assumed to be the magnetic pitch angles. The B-field modes cannot be estimated.

\textit{Three-dimensional axisymmetric spiral B-fields}: This approach uses a three-dimensional regular B-field model with an axisymmetric spiral B-field configuration. This B-field model is a mode of a galactic dynamo with a symmetric spiral pattern in the galactic midplane with a helical component. This method has been used in the FIR polarimetric observations of Centaurus A \citep{ELR2021}, radio polarimetric observations of other galaxies \citep{Braun2010}, and in our Galaxy \citep{RG2010}. The advantages of this method are that the three-dimensional B-field component, and the  pixel-by-pixel pitch angles across the disk can be obtained. The disadvantages are that a parametric B-field model has to be assumed, and the best-fit B-field is not unique due to the large number of free parameters and the ambiguity of the three-dimensional information of the B-fields from observations \citep[e.g.,][]{Braun2010,RG2010}. In addition, the B-field modes cannot be estimated.

%%%%%%%%%%%%%%%%%%
%%%% FIGURE 7 %%%%
%%%%%%%%%%%%%%%%%%
\begin{figure}[ht!]
\includegraphics[width=\columnwidth]{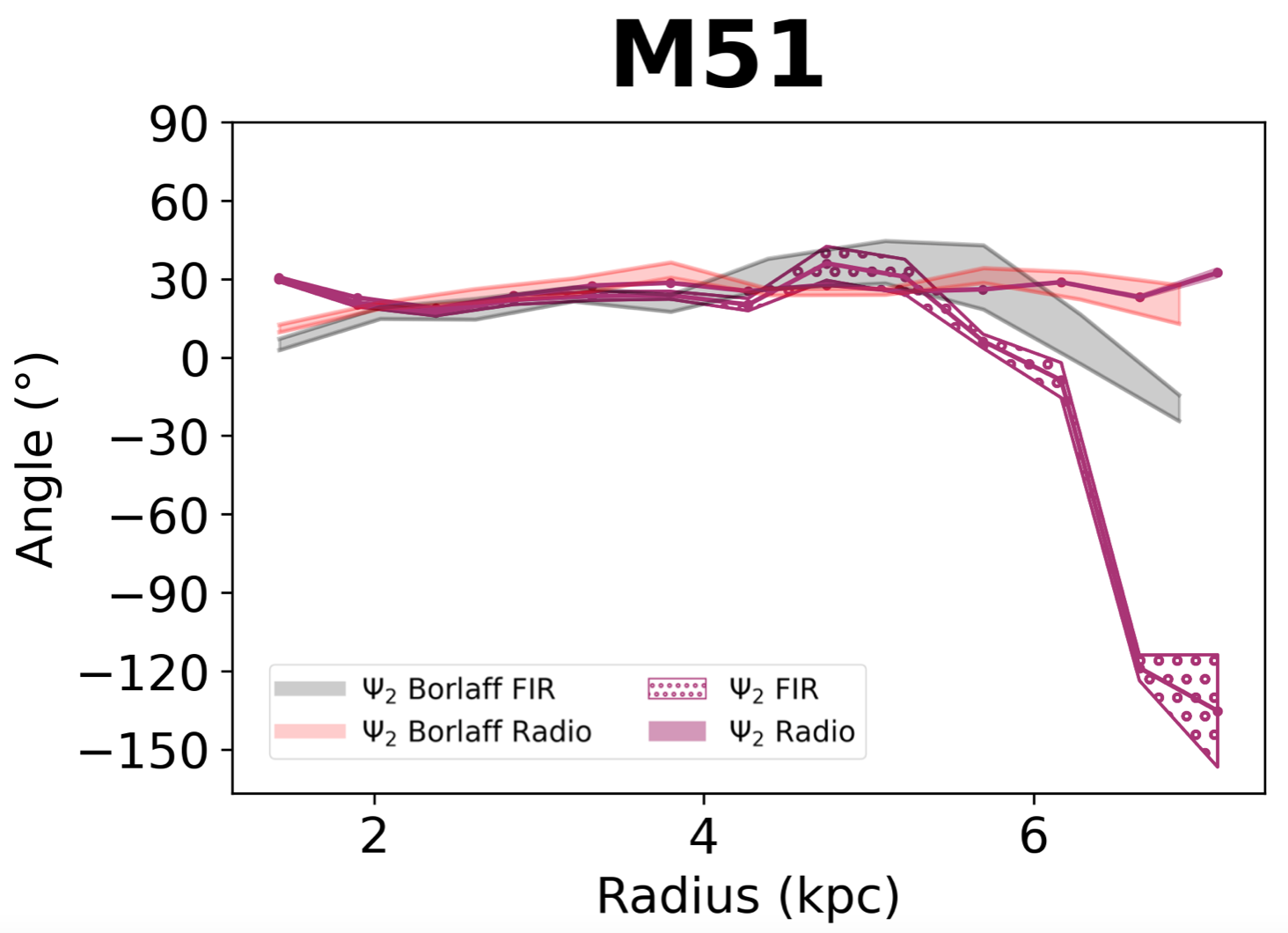}
\caption{Comparison of magnetic pitch angles measurements, $\Psi_2$, between methods at FIR and radio wavelengths. Our method and the pitch angles by \citet{SALSAI} are shown. 
 \label{fig:fig7}}
\end{figure}
%%%%%%%%%%%%%%

The aforementioned models provide the magnetic pitch angle from the measured B-field orientations. Note that our estimated $\Psi_{2}$ is the intrinsic magnetic pitch angle associated with a purely spiral axisymmetric B-field.  Figure \ref{fig:fig7} shows a comparison of the measured magnetic pitch angles using \textsc{mohawc} by \citet{SALSAI} and our pitch angle, $\Psi_{2}$, for M51. \citet{SALSAI} showed that the radio and FIR polarimetric observations do not necessarily trace the same B-field component. This result is clearly detected at galactrocentric distances of $r\ge5$ kpc in \citet{SALSAI}, but only if the spiral arms are analyzed separately from the interarm regions. These authors used a mask to separate the arm and interarm regions based on the total integrated emission (i.e., moment 0) of HI. Using the method presented here, the measured FIR and radio magnetic pitch angles are identical when the full disk was analyzed at once \citep[][Fig. 7]{SALSAI}. Our new method obtains the same result -- a difference in the FIR and radio pitch angles at large radius -- but without the need to mask the data to separate physical components of the disk. Figure \ref{fig:fig7} emphasizes the potential of our method to characterize B-field morphologies in the multi-phase ISM using a model-independent approach that does not require masking to separate different galactic components.

%%%%%%%%%%%%%%%%%%%%%%%%%%%%%%%%%%%%%%%%%%%%%%%%%%%%%%%%
\subsection{Broader applications}\label{subsec:FutureApps}
%%%%%%%%%%%%%%%%%%%%%%%%%%%%%%%%%%%%%%%%%%%%%%%%%%%%%%%%

The method presented here is adapted from \citet{Palumbo2020}. The polarimetric linear decomposition has been applied to the B-field orientation generated by magnetohydrodynamic accretion disk simulations and proposed as a model-independent approach to measuring the accretion state of the M87 black hole observed with the EHT \citep[see also][]{EHT2021_VII,EHT2021_VIII,Emami:2022}. While we adapted this decomposition and applied it to the B-fields of spiral galaxies, our method can also be applied to any vector field where a circle or ellipse is a geometry of particular interest. Apart from galaxies, this method could also be adapted to other ISM morphologies, such as supernova remnants or wind-blown bubbles in star-forming regions \citep[e.g.,][]{Tahani2022}, or radio synchrotron loops \citep[e.g.,][]{Vidal:2015}.   

One straightforward extension of the method presented here would be to quantify the morphology of galaxy structure observed via the total intensity distribution at different wavelengths. One very simple approach to perform this analysis would be to compute the spatial gradient of the emission in order to encode morphological information as a vector field. One could then apply this method directly and compare the results to the magnetic field structure. Similarly, one could quantify the intensity morphology using the Hessian, which measures local curvature in the image plane and thus has been widely used for measuring the orientations of filamentary structures in astrophysical observations \citep[e.g.,][]{PlanckCollaborationXXXII:2016}.

Using this approach to compare galaxy structure to galaxy magnetic field structure opens up intriguing possibilities for morphological insights beyond pitch angle comparisons. Here we can draw further upon analogies to the $E/B$ decomposition of the polarization field that is frequently used to characterize CMB polarization and has recently been widely applied to Galactic emission of diverse physical origins \citep[e.g.][]{Clark:2015, Krachmalnicoff:2018, PlanckCollaboration:2016, PlanckCollaboration:2020}. The correlation between the total intensity and $E$- or $B$-mode polarization field in Galactic dust emission is related to the degree of alignment or misalignment of filamentary density structures and the magnetic field \citep{Huffenberger:2020, Clark:2021, Cukierman:2022}. %The quantification of these correlations in Galactic dust polarization %is clearly extended to the formalism presented here, in order to compute the correlation 
Similar correlations could be computed between various modes of magnetic structure and various tracers of galactic emission structure.

%%%%%%%%%%%%%%%%%%%%%%%%%%%%%%%%%%%%%%%%%%%%%%%%%%%%%%%%
\section{Conclusions}\label{sec:CON}
%%%%%%%%%%%%%%%%%%%%%%%%%%%%%%%%%%%%%%%%%%%%%%%%%%%%%%%%

We have adapted and successfully applied a new model-independent B-field decomposition approach to measure the large-scale ordered B-field orientations associated with five spiral galaxies using FIR and radio polarimetric observations. With radio ($6$ cm) measurements, we found that the B-fields of spiral galaxies were mainly composed of $m = 2$ with additional but subdominant contributions from $m = 0$, $m = 1$, and $m = 3$. With FIR ($154$ \um) measurements, the most dominant modes were $m = 2$ and $m = 0$ with smaller relative contributions from $m = 1$ and $m = 3$. At both radio and FIR wavelengths, the overall contribution of $\tilde{|\beta_{m < 0}|}$ was less than $\tilde{|\beta_{m \ge 0}|}$. NGC~6946 is the extreme case in our sample. In this galaxy, radio measurements still showed $m = 2$ to be dominant, with the rest of the modes contributing roughly the same to the overall B-field orientation. By contrast, in the FIR, no particular mode dominates the B-field structure.  

 We also found that the mean pitch angle of these galaxies is smaller in the FIR data than in the radio, i.e. $\langle |\Psi_{2}^{\rm{FIR}}| \rangle < \langle |\Psi_{2}^{\rm{Radio}}| \rangle$. If this trend holds, the implication is that radio spiral B-fields are more open than FIR spiral B-fields. Overall, we find that $\Psi_2$ increases with increasing radius at radio wavelengths, meaning that the magnetic field structure opens out toward the outskirts of the galaxy. With FIR wavelengths we found greater angular dispersion than with radio wavelengths, indicating that FIR spiral B-fields are less ordered than radio spiral B-fields.

%%%%%%%%%%%%%%%%%%%%%%%%%
%%%% ACKNOWLEDGMENTS %%%%
%%%%%%%%%%%%%%%%%%%%%%%%%

\begin{acknowledgments}
Based on observations made with the NASA/DLR Stratospheric Observatory for Infrared Astronomy (SOFIA) under the  08\_0012 Program. SOFIA is jointly operated by the Universities Space Research Association, Inc. (USRA), under NASA contract NNA17BF53C, and the Deutsches SOFIA Institut (DSI) under DLR contract 50 OK 0901 to the University of Stuttgart. 
\end{acknowledgments}

\appendix 

\section{Amplitudes of high order modes}\label{App:A1}

\begin{figure*}[ht!]
\includegraphics[width=\textwidth]{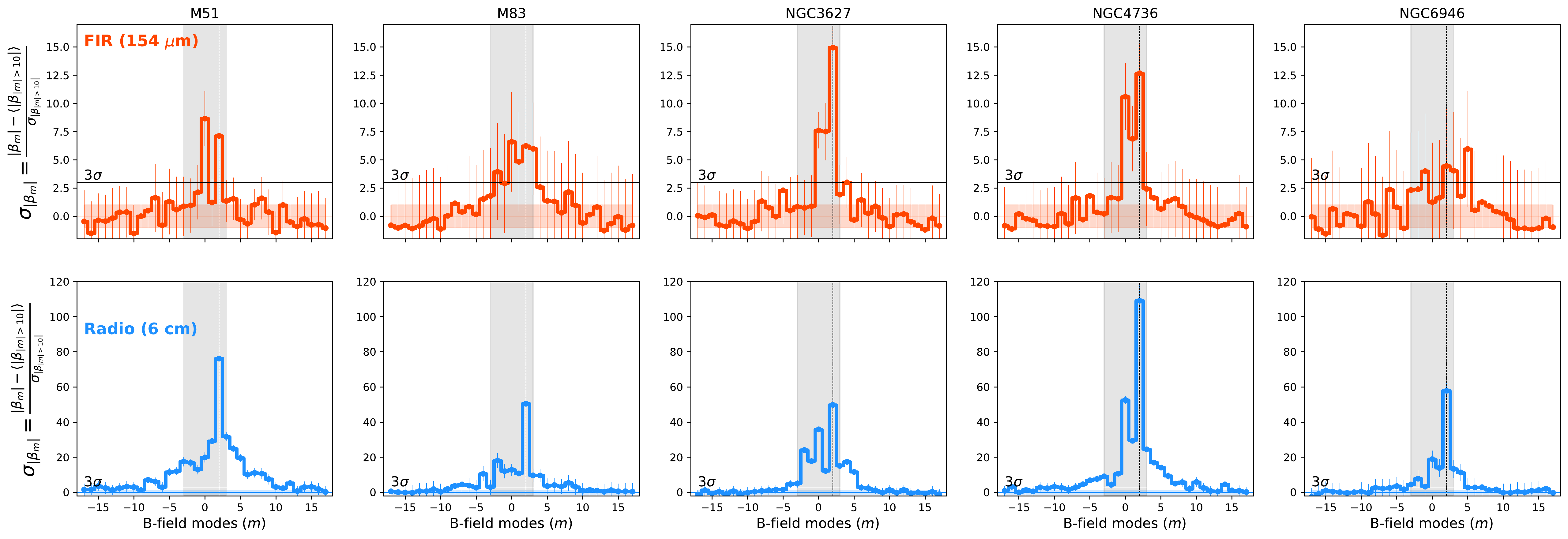}
\caption{Statistical significance of $|\beta_{m}|$ per B-field mode for each galaxy. Each panel shows the $\sigma_{|\beta_{m}|}$ at FIR (top) and radio (bottom) for each mode with their errorbars estimated as the propagated statistical errors of the full galaxy. The $\pm\sigma_{|\beta_{|m| > 10}|}$ (red shadowed region) and the $3\sigma_{|\beta_{|m| > 10}|}$ (black solid line) are shown. For reference, the $m=2$ (black dashed line) and the $m=[-3,3]$ (grey shadowed region) are shown.
 \label{fig:A1}}
\end{figure*}

We estimate the statistical significance of the B-field modes to select the range of modes to be used in this work. We compute the absolute values of the amplitudes, $|\beta_{m}|$, in the $m=[-17,17]$ range using the same methodology as described in Section \ref{subsec:ModelSteps}. The statistical significance of $|\beta_{m}|$ is estimated as

\begin{equation}
\sigma_{|\beta_{m}|} = \frac{|\beta_{m}| - \langle |\beta_{|m| > 10}| \rangle}{\sigma_{|\beta_{|m| > 10}|}},
\end{equation}
\noindent 
where $\langle |\beta_{|m| > 10}|\rangle$ is the median of $|\beta_{m}|$ for $|m|>10$, and $\sigma_{|\beta_{|m| > 10}|}$ is the standard deviation of  $|\beta_{m}|$ for $|m|>10$. Figure \ref{fig:A1} shows $\sigma_{|\beta_{m}|}$ for each mode for the five galaxies studied in this work. 

%%%%%%%%%%%%%%%%%%
%%%% FIGURE 7 %%%%
%%%%%%%%%%%%%%%%%%
%%%%%%%%%%%%%%

The $|m|>10$ was selected because the absolute values of these high modes reach a constant level of power. This constant power level indicates that the observations are not sensitive to these high modes. We use the median of the absolute amplitudes in $|m|>10$, i.e., $\langle |\beta_{|m| > 10}|\rangle$, as the minimum power that the observations are sensitive to. We use the variations of the absolute amplitudes in $|m|>10$, i.e., $\sigma_{|\beta_{|m| > 10}|}$, as the measurement of the power sensitivity of our approach. Thus, we can estimate the statistical significance of the modes above a certain level, i.e., $\sigma_{|\beta_{m}|}$. Figure \ref{fig:A1} shows that the low modes, $m=[-3,3]$, are the most significant, $\sigma_{|\beta_{m}|} > 3\sigma_{|\beta_{|m| > 10}|}$. Our observations seems to not be sensitive to high modes, $|m| > 3$, mainly because the complexity of the B-field morphology is at smaller scales than the resolution of the observations. Thus, we use $m=[-3,3]$ for our study (Section \ref{subsubsec:amplitudes}).

Note that $m=2$ (spiral B-field) is the most prominent B-field mode for most of the galaxies in the FIR and all galaxies in the radio. For M\,83 and NGC\,6946 in the FIR, the absolute amplitudes have large uncertainties indicating that there is not a clear large-scale ordered B-field on these two galaxies. For NGC\,6946, this result is clearly observed in the B-field orientation maps  (Fig. \ref{fig:fig3}). For M\,83, this result seems at odds with the B-field orientation maps. However, note that we are estimating the amplitudes for the full disk. Figure \ref{fig:fig4} shows that the spiral arm dominates  at $r<4$\,kpc and a more complex B-field morphology is present at $r>4$\,kpc.

%% To help institutions obtain information on the effectiveness of their 
%% telescopes the AAS Journals has created a group of keywords for telescope 
%% facilities.
%
%% Following the acknowledgments section, use the following syntax and the
%% \facility{} or \facilities{} macros to list the keywords of facilities used 
%% in the research for the paper.  Each keyword is check against the master 
%% list during copy editing.  Individual instruments can be provided in 
%% parentheses, after the keyword, but they are not verified.

\vspace{5mm}
\facilities{SOFIA (HAWC+), VLA}

%% Similar to \facility{}, there is the optional \software command to allow 
%% authors a place to specify which programs were used during the creation of 
%% the manuscript. Authors should list each code and include either a
%% citation or url to the code inside ()s when available.

\software{\textsc{aplpy} \citep{aplpy2012,aplpy2019},  
          \textsc{astropy} \citep{astropy:2022},
          \textsc{pandas} \citep{pandas},
          \textsc{matplotlib} \citep{matplotlib},
          \textsc{scipy} \citep{scipy}}
%% For this sample we use BibTeX plus aasjournals.bst to generate the
%% the bibliography. The sample631.bib file was populated from ADS. To
%% get the citations to show in the compiled file do the following:
%%
%% pdflatex sample631.tex
%% bibtext sample631
%% pdflatex sample631.tex
%% pdflatex sample631.tex

\bibliography{references}{}

\begin{thebibliography}{}
\expandafter\ifx\csname natexlab\endcsname\relax\def\natexlab#1{#1}\fi
\providecommand{\url}[1]{\href{#1}{#1}}
\providecommand{\dodoi}[1]{doi:~\href{http://doi.org/#1}{\nolinkurl{#1}}}
\providecommand{\doeprint}[1]{\href{http://ascl.net/#1}{\nolinkurl{http://ascl.net/#1}}}
\providecommand{\doarXiv}[1]{\href{https://arxiv.org/abs/#1}{\nolinkurl{https://arxiv.org/abs/#1}}}

\bibitem[{{Astropy Collaboration} {et~al.}(2022){Astropy Collaboration},
  {Price-Whelan}, {Lim}, {Earl}, {Starkman}, {Bradley}, {Shupe}, {Patil},
  {Corrales}, {Brasseur}, {N{"o}the}, {Donath}, {Tollerud}, {Morris},
  {Ginsburg}, {Vaher}, {Weaver}, {Tocknell}, {Jamieson}, {van Kerkwijk},
  {Robitaille}, {Merry}, {Bachetti}, {G{"u}nther}, {Aldcroft},
  {Alvarado-Montes}, {Archibald}, {B{'o}di}, {Bapat}, {Barentsen}, {Baz{'a}n},
  {Biswas}, {Boquien}, {Burke}, {Cara}, {Cara}, {Conroy}, {Conseil}, {Craig},
  {Cross}, {Cruz}, {D'Eugenio}, {Dencheva}, {Devillepoix}, {Dietrich},
  {Eigenbrot}, {Erben}, {Ferreira}, {Foreman-Mackey}, {Fox}, {Freij}, {Garg},
  {Geda}, {Glattly}, {Gondhalekar}, {Gordon}, {Grant}, {Greenfield}, {Groener},
  {Guest}, {Gurovich}, {Handberg}, {Hart}, {Hatfield-Dodds}, {Homeier},
  {Hosseinzadeh}, {Jenness}, {Jones}, {Joseph}, {Kalmbach}, {Karamehmetoglu},
  {Ka{l}uszy{'n}ski}, {Kelley}, {Kern}, {Kerzendorf}, {Koch}, {Kulumani},
  {Lee}, {Ly}, {Ma}, {MacBride}, {Maljaars}, {Muna}, {Murphy}, {Norman},
  {O'Steen}, {Oman}, {Pacifici}, {Pascual}, {Pascual-Granado}, {Patil},
  {Perren}, {Pickering}, {Rastogi}, {Roulston}, {Ryan}, {Rykoff}, {Sabater},
  {Sakurikar}, {Salgado}, {Sanghi}, {Saunders}, {Savchenko}, {Schwardt},
  {Seifert-Eckert}, {Shih}, {Jain}, {Shukla}, {Sick}, {Simpson},
  {Singanamalla}, {Singer}, {Singhal}, {Sinha}, {Sip{H{o}}cz}, {Spitler},
  {Stansby}, {Streicher}, {{{S}}umak}, {Swinbank}, {Taranu}, {Tewary},
  {Tremblay}, {Val-Borro}, {Van Kooten}, {Vasovi{'c}}, {Verma}, {de Miranda
  Cardoso}, {Williams}, {Wilson}, {Winkel}, {Wood-Vasey}, {Xue}, {Yoachim},
  {Zhang}, {Zonca}, \& {Astropy Project Contributors}}]{astropy:2022}
{Astropy Collaboration}, {Price-Whelan}, A.~M., {Lim}, P.~L., {et~al.} 2022,
  apj, 935, 167, \dodoi{10.3847/1538-4357/ac7c74}

\bibitem[{{Beck}(1991)}]{Beck1991}
{Beck}, R. 1991, \aap, 251, 15

\bibitem[{{Beck}(2007)}]{Beck2007}
---. 2007, \aap, 470, 539, \dodoi{10.1051/0004-6361:20066988}

\bibitem[{{Beck} {et~al.}(2019){Beck}, {Chamandy}, {Elson}, \&
  {Blackman}}]{Beck2019}
{Beck}, R., {Chamandy}, L., {Elson}, E., \& {Blackman}, E.~G. 2019, Galaxies,
  8, 4, \dodoi{10.3390/galaxies8010004}

\bibitem[{{Berkhuijsen} {et~al.}(1997){Berkhuijsen}, {Horellou}, {Krause},
  {Neininger}, {Poezd}, {Shukurov}, \& {Sokoloff}}]{Berkhuijsen1997}
{Berkhuijsen}, E.~M., {Horellou}, C., {Krause}, M., {et~al.} 1997, \aap, 318,
  700.
\newblock \doarXiv{astro-ph/9610182}

\bibitem[{{Borlaff} {et~al.}(2021){Borlaff}, {Lopez-Rodriguez}, {Beck},
  {Stepanov}, {Ntormousi}, {Hughes}, {Tassis}, {Marcum}, {Grosset}, {Beckman},
  {Proudfit}, {Clark}, {D{\'\i}az-Santos}, {Mao}, {Reach}, {Roman-Duval},
  {Subramanian}, {Tram}, {Zweibel}, {Dale}, \& {Legacy Team}}]{SALSAI}
{Borlaff}, A.~S., {Lopez-Rodriguez}, E., {Beck}, R., {et~al.} 2021, \apj, 921,
  128, \dodoi{10.3847/1538-4357/ac16d7}

\bibitem[{{Borlaff} {et~al.}(2023){Borlaff}, {Lopez-Rodriguez}, {Beck},
  {Clark}, {Ntormousi}, {Tassis}, {Martin-Alvarez}, {Tahani}, {Dale}, {del
  Moral Castro}, {Roman-Duval}, {Marcum}, {Beckman}, \& {Subramanian}}]{SALSAV}
---. 2023, arXiv e-prints, arXiv:2303.13586, \dodoi{10.48550/arXiv.2303.13586}

\bibitem[{{Brandenburg} \& {Subramanian}(2005)}]{BS2005}
{Brandenburg}, A., \& {Subramanian}, K. 2005, \physrep, 417, 1,
  \dodoi{10.1016/j.physrep.2005.06.005}

\bibitem[{{Braun} {et~al.}(2010){Braun}, {Heald}, \& {Beck}}]{Braun2010}
{Braun}, R., {Heald}, G., \& {Beck}, R. 2010, \aap, 514, A42,
  \dodoi{10.1051/0004-6361/200913375}

\bibitem[{{Chy{\.z}y} \& {Buta}(2008)}]{Chyzy2008}
{Chy{\.z}y}, K.~T., \& {Buta}, R.~J. 2008, \apjl, 677, L17,
  \dodoi{10.1086/587958}

\bibitem[{{Clark} {et~al.}(2015){Clark}, {Hill}, {Peek}, {Putman}, \&
  {Babler}}]{Clark:2015}
{Clark}, S.~E., {Hill}, J.~C., {Peek}, J.~E.~G., {Putman}, M.~E., \& {Babler},
  B.~L. 2015, \prl, 115, 241302, \dodoi{10.1103/PhysRevLett.115.241302}

\bibitem[{{Clark} {et~al.}(2021){Clark}, {Kim}, {Hill}, \&
  {Hensley}}]{Clark:2021}
{Clark}, S.~E., {Kim}, C.-G., {Hill}, J.~C., \& {Hensley}, B.~S. 2021, \apj,
  919, 53, \dodoi{10.3847/1538-4357/ac0e35}

\bibitem[{{Colombo} {et~al.}(2014){Colombo}, {Meidt}, {Schinnerer},
  {Garc{\'\i}a-Burillo}, {Hughes}, {Pety}, {Leroy}, {Dobbs}, {Dumas},
  {Thompson}, {Schuster}, \& {Kramer}}]{Colombo2014}
{Colombo}, D., {Meidt}, S.~E., {Schinnerer}, E., {et~al.} 2014, \apj, 784, 4,
  \dodoi{10.1088/0004-637X/784/1/4}

\bibitem[{{Crosthwaite} {et~al.}(2002){Crosthwaite}, {Turner}, {Buchholz},
  {Ho}, \& {Martin}}]{Crosthwaite2002}
{Crosthwaite}, L.~P., {Turner}, J.~L., {Buchholz}, L., {Ho}, P. T.~P., \&
  {Martin}, R.~N. 2002, \aj, 123, 1892, \dodoi{10.1086/339479}

\bibitem[{{Cukierman} {et~al.}(2022){Cukierman}, {Clark}, \&
  {Halal}}]{Cukierman:2022}
{Cukierman}, A.~J., {Clark}, S.~E., \& {Halal}, G. 2022, arXiv e-prints,
  arXiv:2208.07382, \dodoi{10.48550/arXiv.2208.07382}

\bibitem[{{Daigle} {et~al.}(2006){Daigle}, {Carignan}, {Amram}, {Hernandez},
  {Chemin}, {Balkowski}, \& {Kennicutt}}]{Daigle2006}
{Daigle}, O., {Carignan}, C., {Amram}, P., {et~al.} 2006, \mnras, 367, 469,
  \dodoi{10.1111/j.1365-2966.2006.10002.x}

\bibitem[{{Dicaire} {et~al.}(2008){Dicaire}, {Carignan}, {Amram}, {Hernandez},
  {Chemin}, {Daigle}, {de Denus-Baillargeon}, {Balkowski}, {Boselli}, {Fathi},
  \& {Kennicutt}}]{Dicaire2008}
{Dicaire}, I., {Carignan}, C., {Amram}, P., {et~al.} 2008, \mnras, 385, 553,
  \dodoi{10.1111/j.1365-2966.2008.12868.x}

\bibitem[{{Ehle} \& {Beck}(1993)}]{Ehle1993}
{Ehle}, M., \& {Beck}, R. 1993, \aap, 273, 45

\bibitem[{{Emami} {et~al.}(2022){Emami}, {Ricarte}, {Wong}, {Palumbo}, {Chang},
  {Doeleman}, {Broaderick}, {Narayan}, {Weintroub}, {Wielgus}, {Blackburn},
  {Prather}, {Chael}, {Anantua}, {Chatterjee}, {Marti-Vidal}, {Gomez},
  {Akiyama}, {Liska}, {Hernquist}, {Tremblay}, {Vogelsberger}, {Alcock},
  {Smith}, {Steiner}, {Tiede}, \& {Roelofs}}]{Emami:2022}
{Emami}, R., {Ricarte}, A., {Wong}, G.~N., {et~al.} 2022, arXiv e-prints,
  arXiv:2210.01218, \dodoi{10.48550/arXiv.2210.01218}

\bibitem[{{Event Horizon Telescope Collaboration}
  {et~al.}(2021{\natexlab{a}}){Event Horizon Telescope Collaboration},
  {Akiyama}, {Algaba}, {Alberdi}, {Alef}, {Anantua}, {Asada}, {Azulay},
  {Baczko}, {Ball}, {Balokovi{\'c}}, {Barrett}, {Benson}, {Bintley},
  {Blackburn}, {Blundell}, {Boland}, {Bouman}, {Bower}, {Boyce}, {Bremer},
  {Brinkerink}, {Brissenden}, {Britzen}, {Broderick}, {Broguiere}, {Bronzwaer},
  {Byun}, {Carlstrom}, {Chael}, {Chan}, {Chatterjee}, {Chatterjee}, {Chen},
  {Chen}, {Chesler}, {Cho}, {Christian}, {Conway}, {Cordes}, {Crawford},
  {Crew}, {Cruz-Osorio}, {Cui}, {Davelaar}, {De Laurentis}, {Deane}, {Dempsey},
  {Desvignes}, {Dexter}, {Doeleman}, {Eatough}, {Falcke}, {Farah}, {Fish},
  {Fomalont}, {Ford}, {Fraga-Encinas}, {Freeman}, {Friberg}, {Fromm},
  {Fuentes}, {Galison}, {Gammie}, {Garc{\'\i}a}, {Gentaz}, {Georgiev}, {Goddi},
  {Gold}, {G{\'o}mez}, {G{\'o}mez-Ruiz}, {Gu}, {Gurwell}, {Hada}, {Haggard},
  {Hecht}, {Hesper}, {Ho}, {Ho}, {Honma}, {Huang}, {Huang}, {Hughes}, {Ikeda},
  {Inoue}, {Issaoun}, {James}, {Jannuzi}, {Janssen}, {Jeter}, {Jiang},
  {Jimenez-Rosales}, {Johnson}, {Jorstad}, {Jung}, {Karami}, {Karuppusamy},
  {Kawashima}, {Keating}, {Kettenis}, {Kim}, {Kim}, {Kim}, {Kim}, {Kino},
  {Koay}, {Kofuji}, {Koch}, {Koyama}, {Kramer}, {Kramer}, {Krichbaum}, {Kuo},
  {Lauer}, {Lee}, {Levis}, {Li}, {Li}, {Lindqvist}, {Lico}, {Lindahl}, {Liu},
  {Liu}, {Liuzzo}, {Lo}, {Lobanov}, {Loinard}, {Lonsdale}, {Lu}, {MacDonald},
  {Mao}, {Marchili}, {Markoff}, {Marrone}, {Marscher}, {Mart{\'\i}-Vidal},
  {Matsushita}, {Matthews}, {Medeiros}, {Menten}, {Mizuno}, {Mizuno}, {Moran},
  {Moriyama}, {Moscibrodzka}, {M{\"u}ller}, {Musoke}, {Mej{\'\i}as},
  {Michalik}, {Nadolski}, {Nagai}, {Nagar}, {Nakamura}, {Narayan}, {Narayanan},
  {Natarajan}, {Nathanail}, {Neilsen}, {Neri}, {Ni}, {Noutsos}, {Nowak},
  {Okino}, {Olivares}, {Ortiz-Le{\'o}n}, {Oyama}, {{\"O}zel}, {Palumbo},
  {Park}, {Patel}, {Pen}, {Pesce}, {Pi{\'e}tu}, {Plambeck}, {PopStefanija},
  {Porth}, {P{\"o}tzl}, {Prather}, {Preciado-L{\'o}pez}, {Psaltis}, {Pu},
  {Ramakrishnan}, {Rao}, {Rawlings}, {Raymond}, {Rezzolla}, {Ricarte},
  {Ripperda}, {Roelofs}, {Rogers}, {Ros}, {Rose}, {Roshanineshat}, {Rottmann},
  {Roy}, {Ruszczyk}, {Rygl}, {S{\'a}nchez}, {S{\'a}nchez-Arguelles}, {Sasada},
  {Savolainen}, {Schloerb}, {Schuster}, {Shao}, {Shen}, {Small}, {Sohn},
  {SooHoo}, {Sun}, {Tazaki}, {Tetarenko}, {Tiede}, {Tilanus}, {Titus}, {Toma},
  {Torne}, {Trent}, {Traianou}, {Trippe}, {van Bemmel}, {van Langevelde}, {van
  Rossum}, {Wagner}, {Ward-Thompson}, {Wardle}, {Weintroub}, {Wex}, {Wharton},
  {Wielgus}, {Wong}, {Wu}, {Yoon}, {Young}, {Young}, {Younsi}, {Yuan}, {Yuan},
  {Zensus}, {Zhao}, \& {Zhao}}]{EHT2021_VII}
{Event Horizon Telescope Collaboration}, {Akiyama}, K., {Algaba}, J.~C.,
  {et~al.} 2021{\natexlab{a}}, \apjl, 910, L12,
  \dodoi{10.3847/2041-8213/abe71d}

\bibitem[{{Event Horizon Telescope Collaboration}
  {et~al.}(2021{\natexlab{b}}){Event Horizon Telescope Collaboration},
  {Akiyama}, {Algaba}, {Alberdi}, {Alef}, {Anantua}, {Asada}, {Azulay},
  {Baczko}, {Ball}, {Balokovi{\'c}}, {Barrett}, {Benson}, {Bintley},
  {Blackburn}, {Blundell}, {Boland}, {Bouman}, {Bower}, {Boyce}, {Bremer},
  {Brinkerink}, {Brissenden}, {Britzen}, {Broderick}, {Broguiere}, {Bronzwaer},
  {Byun}, {Carlstrom}, {Chael}, {Chan}, {Chatterjee}, {Chatterjee}, {Chen},
  {Chen}, {Chesler}, {Cho}, {Christian}, {Conway}, {Cordes}, {Crawford},
  {Crew}, {Cruz-Osorio}, {Cui}, {Davelaar}, {De Laurentis}, {Deane}, {Dempsey},
  {Desvignes}, {Dexter}, {Doeleman}, {Eatough}, {Falcke}, {Farah}, {Fish},
  {Fomalont}, {Ford}, {Fraga-Encinas}, {Friberg}, {Fromm}, {Fuentes},
  {Galison}, {Gammie}, {Garc{\'\i}a}, {Gelles}, {Gentaz}, {Georgiev}, {Goddi},
  {Gold}, {G{\'o}mez}, {G{\'o}mez-Ruiz}, {Gu}, {Gurwell}, {Hada}, {Haggard},
  {Hecht}, {Hesper}, {Himwich}, {Ho}, {Ho}, {Honma}, {Huang}, {Huang},
  {Hughes}, {Ikeda}, {Inoue}, {Issaoun}, {James}, {Jannuzi}, {Janssen},
  {Jeter}, {Jiang}, {Jimenez-Rosales}, {Johnson}, {Jorstad}, {Jung}, {Karami},
  {Karuppusamy}, {Kawashima}, {Keating}, {Kettenis}, {Kim}, {Kim}, {Kim},
  {Kim}, {Kino}, {Koay}, {Kofuji}, {Koch}, {Koyama}, {Kramer}, {Kramer},
  {Krichbaum}, {Kuo}, {Lauer}, {Lee}, {Levis}, {Li}, {Li}, {Lindqvist}, {Lico},
  {Lindahl}, {Liu}, {Liu}, {Liuzzo}, {Lo}, {Lobanov}, {Loinard}, {Lonsdale},
  {Lu}, {MacDonald}, {Mao}, {Marchili}, {Markoff}, {Marrone}, {Marscher},
  {Mart{\'\i}-Vidal}, {Matsushita}, {Matthews}, {Medeiros}, {Menten}, {Mizuno},
  {Mizuno}, {Moran}, {Moriyama}, {Moscibrodzka}, {M{\"u}ller}, {Musoke}, {Mus
  Mej{\'\i}as}, {Michalik}, {Nadolski}, {Nagai}, {Nagar}, {Nakamura},
  {Narayan}, {Narayanan}, {Natarajan}, {Nathanail}, {Neilsen}, {Neri}, {Ni},
  {Noutsos}, {Nowak}, {Okino}, {Olivares}, {Ortiz-Le{\'o}n}, {Oyama},
  {{\"O}zel}, {Palumbo}, {Park}, {Patel}, {Pen}, {Pesce}, {Pi{\'e}tu},
  {Plambeck}, {PopStefanija}, {Porth}, {P{\"o}tzl}, {Prather},
  {Preciado-L{\'o}pez}, {Psaltis}, {Pu}, {Ramakrishnan}, {Rao}, {Rawlings},
  {Raymond}, {Rezzolla}, {Ricarte}, {Ripperda}, {Roelofs}, {Rogers}, {Ros},
  {Rose}, {Roshanineshat}, {Rottmann}, {Roy}, {Ruszczyk}, {Rygl},
  {S{\'a}nchez}, {S{\'a}nchez-Arguelles}, {Sasada}, {Savolainen}, {Schloerb},
  {Schuster}, {Shao}, {Shen}, {Small}, {Sohn}, {SooHoo}, {Sun}, {Tazaki},
  {Tetarenko}, {Tiede}, {Tilanus}, {Titus}, {Toma}, {Torne}, {Trent},
  {Traianou}, {Trippe}, {van Bemmel}, {van Langevelde}, {van Rossum}, {Wagner},
  {Ward-Thompson}, {Wardle}, {Weintroub}, {Wex}, {Wharton}, {Wielgus}, {Wong},
  {Wu}, {Yoon}, {Young}, {Young}, {Younsi}, {Yuan}, {Yuan}, {Zensus}, {Zhao},
  \& {Zhao}}]{EHT2021_VIII}
---. 2021{\natexlab{b}}, \apjl, 910, L13, \dodoi{10.3847/2041-8213/abe4de}

\bibitem[{{Fletcher} {et~al.}(2011){Fletcher}, {Beck}, {Shukurov},
  {Berkhuijsen}, \& {Horellou}}]{Fletcher2011}
{Fletcher}, A., {Beck}, R., {Shukurov}, A., {Berkhuijsen}, E.~M., \&
  {Horellou}, C. 2011, \mnras, 412, 2396,
  \dodoi{10.1111/j.1365-2966.2010.18065.x}

\bibitem[{{Fletcher} {et~al.}(2004){Fletcher}, {Berkhuijsen}, {Beck}, \&
  {Shukurov}}]{Fletcher2004}
{Fletcher}, A., {Berkhuijsen}, E.~M., {Beck}, R., \& {Shukurov}, A. 2004, \aap,
  414, 53, \dodoi{10.1051/0004-6361:20034133}

\bibitem[{{Frick} {et~al.}(2000){Frick}, {Beck}, {Shukurov}, {Sokoloff},
  {Ehle}, \& {Kamphuis}}]{Frick2000}
{Frick}, P., {Beck}, R., {Shukurov}, A., {et~al.} 2000, \mnras, 318, 925,
  \dodoi{10.1046/j.1365-8711.2000.03783.x}

\bibitem[{{Frick} {et~al.}(2016){Frick}, {Stepanov}, {Beck}, {Sokoloff},
  {Shukurov}, {Ehle}, \& {Lundgren}}]{Frick2016}
{Frick}, P., {Stepanov}, R., {Beck}, R., {et~al.} 2016, \aap, 585, A21,
  \dodoi{10.1051/0004-6361/201526796}

\bibitem[{{Haverkorn}(2015)}]{Haverkorn:2015}
{Haverkorn}, M. 2015, in Astrophysics and Space Science Library, Vol. 407,
  Magnetic Fields in Diffuse Media, ed. A.~{Lazarian}, E.~M. {de Gouveia Dal
  Pino}, \& C.~{Melioli}, 483, \dodoi{10.1007/978-3-662-44625-6_17}

\bibitem[{{Haverkorn} {et~al.}(2008){Haverkorn}, {Brown}, {Gaensler}, \&
  {McClure-Griffiths}}]{Haverkorn2008}
{Haverkorn}, M., {Brown}, J.~C., {Gaensler}, B.~M., \& {McClure-Griffiths},
  N.~M. 2008, \apj, 680, 362, \dodoi{10.1086/587165}

\bibitem[{{Huffenberger} {et~al.}(2020){Huffenberger}, {Rotti}, \&
  {Collins}}]{Huffenberger:2020}
{Huffenberger}, K.~M., {Rotti}, A., \& {Collins}, D.~C. 2020, \apj, 899, 31,
  \dodoi{10.3847/1538-4357/ab9df9}

\bibitem[{Hunter(2007)}]{matplotlib}
Hunter, J.~D. 2007, Computing in Science \& Engineering, 9, 90,
  \dodoi{10.1109/MCSE.2007.55}

\bibitem[{{Kamionkowski} {et~al.}(1997){Kamionkowski}, {Kosowsky}, \&
  {Stebbins}}]{Kamionkowski1997}
{Kamionkowski}, M., {Kosowsky}, A., \& {Stebbins}, A. 1997, \prl, 78, 2058,
  \dodoi{10.1103/PhysRevLett.78.2058}

\bibitem[{{Karachentsev} {et~al.}(2000){Karachentsev}, {Sharina}, \&
  {Huchtmeier}}]{Karachentsev2000}
{Karachentsev}, I.~D., {Sharina}, M.~E., \& {Huchtmeier}, W.~K. 2000, \aap,
  362, 544.
\newblock \doarXiv{astro-ph/0010148}

\bibitem[{{Kenney} \& {Lord}(1991)}]{Kenney1991}
{Kenney}, J. D.~P., \& {Lord}, S.~D. 1991, \apj, 381, 118,
  \dodoi{10.1086/170634}

\bibitem[{{Kennicutt} {et~al.}(2003){Kennicutt}, {Armus}, {Bendo}, {Calzetti},
  {Dale}, {Draine}, {Engelbracht}, {Gordon}, {Grauer}, {Helou}, {Hollenbach},
  {Jarrett}, {Kewley}, {Leitherer}, {Li}, {Malhotra}, {Regan}, {Rieke},
  {Rieke}, {Roussel}, {Smith}, {Thornley}, \& {Walter}}]{Kennicutt2003}
{Kennicutt}, Robert~C., J., {Armus}, L., {Bendo}, G., {et~al.} 2003, \pasp,
  115, 928, \dodoi{10.1086/376941}

\bibitem[{{Krachmalnicoff} {et~al.}(2018){Krachmalnicoff}, {Carretti},
  {Baccigalupi}, {Bernardi}, {Brown}, {Gaensler}, {Haverkorn}, {Kesteven},
  {Perrotta}, {Poppi}, \& {Staveley-Smith}}]{Krachmalnicoff:2018}
{Krachmalnicoff}, N., {Carretti}, E., {Baccigalupi}, C., {et~al.} 2018, \aap,
  618, A166, \dodoi{10.1051/0004-6361/201832768}

\bibitem[{{Krasheninnikova} {et~al.}(1989){Krasheninnikova}, {Shukurov},
  {Ruzmaikin}, \& {Sokolov}}]{Krasheninnikova1989}
{Krasheninnikova}, I., {Shukurov}, A., {Ruzmaikin}, A., \& {Sokolov}, D. 1989,
  \aap, 213, 19

\bibitem[{{Krause} \& {Wielebinski}(1991)}]{KW1991}
{Krause}, F., \& {Wielebinski}, R. 1991, Reviews in Modern Astronomy, 4, 260,
  \dodoi{10.1007/978-3-642-76750-0_18}

\bibitem[{{Krause} {et~al.}(2020){Krause}, {Irwin}, {Schmidt}, {Stein},
  {Miskolczi}, {Carolina Mora-Partiarroyo}, {Wiegert}, {Beck}, {Stil}, {Heald},
  {Li}, {Damas-Segovia}, {Vargas}, {Rand}, {West}, {Walterbos}, {Dettmar},
  {English}, \& {Woodfinden}}]{Krause2020}
{Krause}, M., {Irwin}, J., {Schmidt}, P., {et~al.} 2020, \aap, 639, A112,
  \dodoi{10.1051/0004-6361/202037780}

\bibitem[{{Kuno} {et~al.}(2007){Kuno}, {Sato}, {Nakanishi}, {Hirota}, {Tosaki},
  {Shioya}, {Sorai}, {Nakai}, {Nishiyama}, \& {Vila-Vilar{\'o}}}]{Kuno2007}
{Kuno}, N., {Sato}, N., {Nakanishi}, H., {et~al.} 2007, \pasj, 59, 117,
  \dodoi{10.1093/pasj/59.1.117}

\bibitem[{{Lopez-Rodriguez}(2021)}]{ELR2021}
{Lopez-Rodriguez}, E. 2021, Nature Astronomy, 5, 604,
  \dodoi{10.1038/s41550-021-01329-9}

\bibitem[{{Lopez-Rodriguez} {et~al.}(2021){Lopez-Rodriguez}, {Beck}, {Clark},
  {Hughes}, {Borlaff}, {Ntormousi}, {Grosset}, {Tassis}, {Beckman},
  {Subramanian}, {Dale}, \& {D{\'\i}az-Santos}}]{SALSAII}
{Lopez-Rodriguez}, E., {Beck}, R., {Clark}, S.~E., {et~al.} 2021, \apj, 923,
  150, \dodoi{10.3847/1538-4357/ac2e01}

\bibitem[{{Lopez-Rodriguez} {et~al.}(2022{\natexlab{a}}){Lopez-Rodriguez},
  {Borlaff}, {Beck}, {Reach}, {Mao}, {Ntormousi}, {Tassis}, {Martin-Alvarez},
  {Clark}, {Dale}, \& {del Moral-Castro}}]{SALSAVI}
{Lopez-Rodriguez}, E., {Borlaff}, A.~S., {Beck}, R., {et~al.}
  2022{\natexlab{a}}, arXiv e-prints, arXiv:2211.00012.
\newblock \doarXiv{2211.00012}

\bibitem[{{Lopez-Rodriguez} {et~al.}(2022{\natexlab{b}}){Lopez-Rodriguez},
  {Mao}, {Beck}, {Borlaff}, {Ntormousi}, {Tassis}, {Dale}, {Roman-Duval},
  {Subramanian}, {Martin-Alvarez}, {Marcum}, {Clark}, {Reach}, {Harper}, \&
  {Zweibel}}]{SALSAIV}
{Lopez-Rodriguez}, E., {Mao}, S.~A., {Beck}, R., {et~al.} 2022{\natexlab{b}},
  arXiv e-prints, arXiv:2205.01105.
\newblock \doarXiv{2205.01105}

\bibitem[{{Lopez-Rodriguez} {et~al.}(2022{\natexlab{c}}){Lopez-Rodriguez},
  {Clarke}, {Shenoy}, {Vacca}, {Coude}, {Arneson}, {Ashton}, {Eftekharzadeh},
  {Beck}, {Beckman}, {Borlaff}, {Clark}, {Dale}, {Martin-Alvarez}, {Ntormousi},
  {Reach}, {Roman-Duval}, {Tassis}, {Harper}, \& {Marcum}}]{SALSAIII}
{Lopez-Rodriguez}, E., {Clarke}, M., {Shenoy}, S., {et~al.} 2022{\natexlab{c}},
  arXiv e-prints, arXiv:2204.13611.
\newblock \doarXiv{2204.13611}

\bibitem[{{McQuinn} {et~al.}(2017){McQuinn}, {Skillman}, {Dolphin}, {Berg}, \&
  {Kennicutt}}]{McQuinn2017}
{McQuinn}, K. B.~W., {Skillman}, E.~D., {Dolphin}, A.~E., {Berg}, D., \&
  {Kennicutt}, R. 2017, \aj, 154, 51, \dodoi{10.3847/1538-3881/aa7aad}

\bibitem[{{Palumbo} {et~al.}(2020){Palumbo}, {Wong}, \&
  {Prather}}]{Palumbo2020}
{Palumbo}, D. C.~M., {Wong}, G.~N., \& {Prather}, B.~S. 2020, \apj, 894, 156,
  \dodoi{10.3847/1538-4357/ab86ac}

\bibitem[{pandas~development team(2020)}]{pandas}
pandas~development team, T. 2020, pandas-dev/pandas: Pandas, latest,  Zenodo,
  \dodoi{10.5281/zenodo.3509134}

\bibitem[{{Patrikeev} {et~al.}(2006){Patrikeev}, {Fletcher}, {Stepanov},
  {Beck}, {Berkhuijsen}, {Frick}, \& {Horellou}}]{Patrikeev2006}
{Patrikeev}, I., {Fletcher}, A., {Stepanov}, R., {et~al.} 2006, \aap, 458, 441,
  \dodoi{10.1051/0004-6361:20065225}

\bibitem[{{Planck Collaboration} {et~al.}(2016{\natexlab{a}}){Planck
  Collaboration}, {Adam}, {Ade}, {Aghanim}, {Alves}, {Arnaud}, {Arzoumanian},
  {Ashdown}, {Aumont}, {Baccigalupi}, {Banday}, {Barreiro}, {Bartolo},
  {Battaner}, {Benabed}, {Benoit-L{\'e}vy}, {Bernard}, {Bersanelli},
  {Bielewicz}, {Bonaldi}, {Bonavera}, {Bond}, {Borrill}, {Bouchet},
  {Boulanger}, {Bracco}, {Burigana}, {Butler}, {Calabrese}, {Cardoso},
  {Catalano}, {Chamballu}, {Chiang}, {Christensen}, {Colombi}, {Colombo},
  {Combet}, {Couchot}, {Crill}, {Curto}, {Cuttaia}, {Danese}, {Davies},
  {Davis}, {de Bernardis}, {de Rosa}, {de Zotti}, {Delabrouille}, {Dickinson},
  {Diego}, {Dole}, {Donzelli}, {Dor{\'e}}, {Douspis}, {Ducout}, {Dupac},
  {Efstathiou}, {Elsner}, {En{\ss}lin}, {Eriksen}, {Falgarone}, {Ferri{\`e}re},
  {Finelli}, {Forni}, {Frailis}, {Fraisse}, {Franceschi}, {Frejsel},
  {Galeotta}, {Galli}, {Ganga}, {Ghosh}, {Giard}, {Gjerl{\o}w},
  {Gonz{\'a}lez-Nuevo}, {G{\'o}rski}, {Gregorio}, {Gruppuso}, {Guillet},
  {Hansen}, {Hanson}, {Harrison}, {Henrot-Versill{\'e}},
  {Hern{\'a}ndez-Monteagudo}, {Herranz}, {Hildebrandt}, {Hivon}, {Hobson},
  {Holmes}, {Hovest}, {Huffenberger}, {Hurier}, {Jaffe}, {Jaffe}, {Jones},
  {Juvela}, {Keih{\"a}nen}, {Keskitalo}, {Kisner}, {Kneissl}, {Knoche}, {Kunz},
  {Kurki-Suonio}, {Lagache}, {Lamarre}, {Lasenby}, {Lattanzi}, {Lawrence},
  {Leonardi}, {Levrier}, {Liguori}, {Lilje}, {Linden-V{\o}rnle},
  {L{\'o}pez-Caniego}, {Lubin}, {Mac{\'\i}as-P{\'e}rez}, {Maffei}, {Maino},
  {Mandolesi}, {Maris}, {Marshall}, {Martin}, {Mart{\'\i}nez-Gonz{\'a}lez},
  {Masi}, {Matarrese}, {Mazzotta}, {Melchiorri}, {Mendes}, {Mennella},
  {Migliaccio}, {Miville-Desch{\^e}nes}, {Moneti}, {Montier}, {Morgante},
  {Mortlock}, {Munshi}, {Murphy}, {Naselsky}, {Natoli}, {N{\o}rgaard-Nielsen},
  {Noviello}, {Novikov}, {Novikov}, {Oppermann}, {Oxborrow}, {Pagano}, {Pajot},
  {Paoletti}, {Pasian}, {Perdereau}, {Perotto}, {Perrotta}, {Pettorino},
  {Piacentini}, {Piat}, {Plaszczynski}, {Pointecouteau}, {Polenta}, {Ponthieu},
  {Popa}, {Pratt}, {Prunet}, {Puget}, {Rachen}, {Reach}, {Reinecke},
  {Remazeilles}, {Renault}, {Ristorcelli}, {Rocha}, {Roudier},
  {Rubi{\~n}o-Mart{\'\i}n}, {Rusholme}, {Sandri}, {Santos}, {Savini}, {Scott},
  {Soler}, {Spencer}, {Stolyarov}, {Sudiwala}, {Sunyaev}, {Sutton},
  {Suur-Uski}, {Sygnet}, {Tauber}, {Terenzi}, {Toffolatti}, {Tomasi},
  {Tristram}, {Tucci}, {Umana}, {Valenziano}, {Valiviita}, {Van Tent},
  {Vielva}, {Villa}, {Wade}, {Wandelt}, {Wehus}, {Wiesemeyer}, {Yvon},
  {Zacchei}, \& {Zonca}}]{PlanckCollaborationXXXII:2016}
{Planck Collaboration}, {Adam}, R., {Ade}, P.~A.~R., {et~al.}
  2016{\natexlab{a}}, \aap, 586, A135, \dodoi{10.1051/0004-6361/201425044}

\bibitem[{{Planck Collaboration} {et~al.}(2016{\natexlab{b}}){Planck
  Collaboration}, {Ade}, {Aghanim}, {Arnaud}, {Ashdown}, {Aumont},
  {Baccigalupi}, {Banday}, {Barreiro}, {Bartolo}, {Battaner}, {Benabed},
  {Benoit-L{\'e}vy}, {Bernard}, {Bersanelli}, {Bielewicz}, {Bonaldi},
  {Bonavera}, {Bond}, {Borrill}, {Bouchet}, {Boulanger}, {Bracco}, {Burigana},
  {Calabrese}, {Cardoso}, {Catalano}, {Chamballu}, {Chary}, {Chiang},
  {Christensen}, {Colombo}, {Combet}, {Crill}, {Curto}, {Cuttaia}, {Danese},
  {Davies}, {Davis}, {de Bernardis}, {de Rosa}, {de Zotti}, {Delabrouille},
  {Delouis}, {Dickinson}, {Diego}, {Dole}, {Donzelli}, {Dor{\'e}}, {Douspis},
  {Dunkley}, {Dupac}, {Efstathiou}, {Elsner}, {En{\ss}lin}, {Eriksen},
  {Falgarone}, {Ferri{\`e}re}, {Finelli}, {Forni}, {Frailis}, {Fraisse},
  {Franceschi}, {Frolov}, {Galeotta}, {Galli}, {Ganga}, {Ghosh}, {Giard},
  {Gjerl{\o}w}, {Gonz{\'a}lez-Nuevo}, {G{\'o}rski}, {Gruppuso}, {Guillet},
  {Hansen}, {Harrison}, {Helou}, {Hern{\'a}ndez-Monteagudo}, {Herranz},
  {Hildebrandt}, {Hivon}, {Hornstrup}, {Hovest}, {Huang}, {Huffenberger},
  {Hurier}, {Jaffe}, {Jones}, {Juvela}, {Keih{\"a}nen}, {Keskitalo}, {Kisner},
  {Kneissl}, {Knoche}, {Kunz}, {Kurki-Suonio}, {Lamarre}, {Lasenby},
  {Lattanzi}, {Lawrence}, {Leonardi}, {Le{\'o}n-Tavares}, {Levrier}, {Liguori},
  {Lilje}, {Linden-V{\o}rnle}, {L{\'o}pez-Caniego}, {Lubin},
  {Mac{\'\i}as-P{\'e}rez}, {Maffei}, {Maino}, {Mandolesi}, {Maris}, {Martin},
  {Mart{\'\i}nez-Gonz{\'a}lez}, {Masi}, {Matarrese}, {McGehee}, {Melchiorri},
  {Mennella}, {Migliaccio}, {Miville-Desch{\^e}nes}, {Moneti}, {Montier},
  {Morgante}, {Mortlock}, {Munshi}, {Murphy}, {Naselsky}, {Nati}, {Natoli},
  {Novikov}, {Novikov}, {Oppermann}, {Oxborrow}, {Pagano}, {Pajot}, {Paoletti},
  {Pasian}, {Perdereau}, {Pettorino}, {Piacentini}, {Piat}, {Pierpaoli},
  {Plaszczynski}, {Pointecouteau}, {Polenta}, {Ponthieu}, {Pratt}, {Prunet},
  {Puget}, {Rachen}, {Reach}, {Rebolo}, {Reinecke}, {Remazeilles}, {Renault},
  {Renzi}, {Ristorcelli}, {Rocha}, {Rosset}, {Rossetti}, {Roudier},
  {Rubi{\~n}o-Mart{\'\i}n}, {Rusholme}, {Sandri}, {Santos}, {Savelainen},
  {Savini}, {Scott}, {Serra}, {Soler}, {Stolyarov}, {Sudiwala}, {Sunyaev},
  {Suur-Uski}, {Sygnet}, {Tauber}, {Terenzi}, {Toffolatti}, {Tomasi},
  {Tristram}, {Tucci}, {Umana}, {Valenziano}, {Valiviita}, {Van Tent},
  {Vielva}, {Villa}, {Wade}, {Wandelt}, {Wehus}, {Yvon}, {Zacchei}, \&
  {Zonca}}]{PlanckCollaboration:2016}
{Planck Collaboration}, {Ade}, P.~A.~R., {Aghanim}, N., {et~al.}
  2016{\natexlab{b}}, \aap, 586, A141, \dodoi{10.1051/0004-6361/201526506}

\bibitem[{{Planck Collaboration} {et~al.}(2020){Planck Collaboration},
  {Akrami}, {Ashdown}, {Aumont}, {Baccigalupi}, {Ballardini}, {Banday},
  {Barreiro}, {Bartolo}, {Basak}, {Benabed}, {Bernard}, {Bersanelli},
  {Bielewicz}, {Bond}, {Borrill}, {Bouchet}, {Boulanger}, {Bracco}, {Bucher},
  {Burigana}, {Calabrese}, {Cardoso}, {Carron}, {Chiang}, {Combet}, {Crill},
  {de Bernardis}, {de Zotti}, {Delabrouille}, {Delouis}, {Di Valentino},
  {Dickinson}, {Diego}, {Ducout}, {Dupac}, {Efstathiou}, {Elsner},
  {En{\ss}lin}, {Falgarone}, {Fantaye}, {Ferri{\`e}re}, {Finelli},
  {Forastieri}, {Frailis}, {Fraisse}, {Franceschi}, {Frolov}, {Galeotta},
  {Galli}, {Ganga}, {G{\'e}nova-Santos}, {Ghosh}, {Gonz{\'a}lez-Nuevo},
  {G{\'o}rski}, {Gruppuso}, {Gudmundsson}, {Guillet}, {Handley}, {Hansen},
  {Herranz}, {Huang}, {Jaffe}, {Jones}, {Keih{\"a}nen}, {Keskitalo}, {Kiiveri},
  {Kim}, {Krachmalnicoff}, {Kunz}, {Kurki-Suonio}, {Lamarre}, {Lasenby}, {Le
  Jeune}, {Levrier}, {Liguori}, {Lilje}, {Lindholm}, {L{\'o}pez-Caniego},
  {Lubin}, {Ma}, {Mac{\'\i}as-P{\'e}rez}, {Maggio}, {Maino}, {Mandolesi},
  {Mangilli}, {Martin}, {Mart{\'\i}nez-Gonz{\'a}lez}, {Matarrese}, {McEwen},
  {Meinhold}, {Melchiorri}, {Migliaccio}, {Miville-Desch{\^e}nes}, {Molinari},
  {Moneti}, {Montier}, {Morgante}, {Natoli}, {Pagano}, {Paoletti}, {Pettorino},
  {Piacentini}, {Polenta}, {Puget}, {Rachen}, {Reinecke}, {Remazeilles},
  {Renzi}, {Rocha}, {Rosset}, {Roudier}, {Rubi{\~n}o-Mart{\'\i}n},
  {Ruiz-Granados}, {Salvati}, {Sandri}, {Savelainen}, {Scott}, {Soler},
  {Spencer}, {Tauber}, {Tavagnacco}, {Toffolatti}, {Tomasi}, {Trombetti},
  {Valiviita}, {Vansyngel}, {Van Tent}, {Vielva}, {Villa}, {Vittorio}, {Wehus},
  {Zacchei}, \& {Zonca}}]{PlanckCollaboration:2020}
{Planck Collaboration}, {Akrami}, Y., {Ashdown}, M., {et~al.} 2020, \aap, 641,
  A11, \dodoi{10.1051/0004-6361/201832618}

\bibitem[{Robitaille(2019)}]{aplpy2019}
Robitaille, T. 2019, {APLpy v2.0: The Astronomical Plotting Library in Python},
  \dodoi{10.5281/zenodo.2567476}

\bibitem[{{Robitaille} \& {Bressert}(2012)}]{aplpy2012}
{Robitaille}, T., \& {Bressert}, E. 2012, {APLpy: Astronomical Plotting Library
  in Python}, Astrophysics Source Code Library.
\newblock \doeprint{1208.017}

\bibitem[{{Rohde} {et~al.}(1999){Rohde}, {Beck}, \& {Elstner}}]{Rohde1999}
{Rohde}, R., {Beck}, R., \& {Elstner}, D. 1999, \aap, 350, 423

\bibitem[{{Ruiz-Granados} {et~al.}(2010){Ruiz-Granados},
  {Rubi{\~n}o-Mart{\'\i}n}, \& {Battaner}}]{RG2010}
{Ruiz-Granados}, B., {Rubi{\~n}o-Mart{\'\i}n}, J.~A., \& {Battaner}, E. 2010,
  \aap, 522, A73, \dodoi{10.1051/0004-6361/200912733}

\bibitem[{{Ruzmaikin} {et~al.}(1988){Ruzmaikin}, {Sokolov}, \&
  {Shukurov}}]{Ruzmaikin1988}
{Ruzmaikin}, A., {Sokolov}, D., \& {Shukurov}, A. 1988, \nat, 336, 341,
  \dodoi{10.1038/336341a0}

\bibitem[{{Seljak} \& {Zaldarriaga}(1997)}]{SK1997}
{Seljak}, U., \& {Zaldarriaga}, M. 1997, \prl, 78, 2054,
  \dodoi{10.1103/PhysRevLett.78.2054}

\bibitem[{{Shukurov} \& {Subramanian}(2021)}]{ss21}
{Shukurov}, A., \& {Subramanian}, K. 2021, Astrophysical Magnetic Fields: From
  Galaxies to the Early Universe (Cambridge: Cambridge University Press)

\bibitem[{{Soida} {et~al.}(2001){Soida}, {Urbanik}, {Beck}, {Wielebinski}, \&
  {Balkowski}}]{Soida2001}
{Soida}, M., {Urbanik}, M., {Beck}, R., {Wielebinski}, R., \& {Balkowski}, C.
  2001, \aap, 378, 40, \dodoi{10.1051/0004-6361:20011185}

\bibitem[{{Subramanian}(1998)}]{Subramanian1998}
{Subramanian}, K. 1998, \mnras, 294, 718,
  \dodoi{10.1046/j.1365-8711.1998.01284.x}

\bibitem[{{Tahani} {et~al.}(2022){Tahani}, {Bastien}, {Furuya}, {Pattle},
  {Johnstone}, {Arzoumanian}, {Doi}, {Hasegawa}, {Inutsuka}, {Coud{\'e}},
  {Fissel}, {Chen}, {Poidevin}, {Sadavoy}, {Friesen}, {Koch}, {Di Francesco},
  {Moriarty-Schieven}, {Chen}, {Chung}, {Eswaraiah}, {Fanciullo}, {Gledhill},
  {Le Gouellec}, {Hoang}, {Hwang}, {Kang}, {Kim}, {Kirchschlager}, {Kwon},
  {Lee}, {Liu}, {Onaka}, {Rawlings}, {Soam}, {Tamura}, {Tang}, {Tomisaka},
  {Whitworth}, {Kwon}, {Hoang}, {Redman}, {Berry}, {Ching}, {Wang}, {Lai},
  {Qiu}, {Ward-Thompson}, {Houde}, {Byun}, {Chen}, {Chen}, {Cho}, {Choi},
  {Choi}, {Chrysostomou}, {Diep}, {Duan}, {Fiege}, {Franzmann}, {Friberg},
  {Fuller}, {Graves}, {Greaves}, {Griffin}, {Gu}, {Han}, {Hatchell}, {Hayashi},
  {Hull}, {Inoue}, {Iwasaki}, {Jeong}, {Kanamori}, {Kang}, {Kang}, {Kataoka},
  {Kawabata}, {Kemper}, {Kim}, {Kim}, {Kim}, {Kim}, {Kim}, {Kirk}, {Kobayashi},
  {Konyves}, {Kusune}, {Lacaille}, {Law}, {Lee}, {Lee}, {Lee}, {Lee}, {Lee},
  {Li}, {Li}, {Li}, {Liu}, {Liu}, {Liu}, {de Looze}, {Lyo}, {Mairs},
  {Matsumura}, {Matthews}, {Nagata}, {Nakamura}, {Nakanishi}, {Ohashi}, {Park},
  {Parsons}, {Peretto}, {Pyo}, {Qian}, {Rao}, {Retter}, {Richer}, {Rigby},
  {Saito}, {Savini}, {Scaife}, {Seta}, {Shimajiri}, {Shinnaga}, {Tang},
  {Tsukamoto}, {Viti}, {Wang}, {Yen}, {Yoo}, {Yuan}, {Yun}, {Zenko}, {Zhang},
  {Zhang}, {Zhang}, {Zhou}, {Zhu}, {Andr{\'e}}, {Dowell}, {Eyres}, {Falle},
  {van Loo}, \& {Robitaille}}]{Tahani2022}
{Tahani}, M., {Bastien}, P., {Furuya}, R.~S., {et~al.} 2022, arXiv e-prints,
  arXiv:2212.10884.
\newblock \doarXiv{2212.10884}

\bibitem[{{Tully} {et~al.}(2013){Tully}, {Courtois}, {Dolphin}, {Fisher},
  {H{\'e}raudeau}, {Jacobs}, {Karachentsev}, {Makarov}, {Makarova},
  {Mitronova}, {Rizzi}, {Shaya}, {Sorce}, \& {Wu}}]{Tully2013}
{Tully}, R.~B., {Courtois}, H.~M., {Dolphin}, A.~E., {et~al.} 2013, \aj, 146,
  86, \dodoi{10.1088/0004-6256/146/4/86}

\bibitem[{{Van Eck} {et~al.}(2015){Van Eck}, {Brown}, {Shukurov}, \&
  {Fletcher}}]{VanEck2015}
{Van Eck}, C.~L., {Brown}, J.~C., {Shukurov}, A., \& {Fletcher}, A. 2015, \apj,
  799, 35, \dodoi{10.1088/0004-637X/799/1/35}

\bibitem[{{Vidal} {et~al.}(2015){Vidal}, {Dickinson}, {Davies}, \&
  {Leahy}}]{Vidal:2015}
{Vidal}, M., {Dickinson}, C., {Davies}, R.~D., \& {Leahy}, J.~P. 2015, \mnras,
  452, 656, \dodoi{10.1093/mnras/stv1328}

\bibitem[{Virtanen {et~al.}(2020)Virtanen, Gommers, Oliphant, Haberland, Reddy,
  Cournapeau, Burovski, Peterson, Weckesser, Bright, {van der Walt}, Brett,
  Wilson, Millman, Mayorov, Nelson, Jones, Kern, Larson, Carey, Polat, Feng,
  Moore, {VanderPlas}, Laxalde, Perktold, Cimrman, Henriksen, Quintero, Harris,
  Archibald, Ribeiro, Pedregosa, {van Mulbregt}, \& {SciPy 1.0
  Contributors}}]{scipy}
Virtanen, P., Gommers, R., Oliphant, T.~E., {et~al.} 2020, Nature Methods, 17,
  261, \dodoi{10.1038/s41592-019-0686-2}

\bibitem[{{Zaldarriaga}(2001)}]{Zaldarriaga2001}
{Zaldarriaga}, M. 2001, \prd, 64, 103001, \dodoi{10.1103/PhysRevD.64.103001}

\end{thebibliography}
\bibliographystyle{aasjournal}

%% This command is needed to show the entire author+affiliation list when
%% the collaboration and author truncation commands are used.  It has to
%% go at the end of the manuscript.
%\allauthors

%% Include this line if you are using the \added, \replaced, \deleted
%% commands to see a summary list of all changes at the end of the article.
%\listofchanges

\end{document}